\documentclass{article}

\usepackage[style=alphabetic,maxbibnames=99,maxalphanames=3,natbib=true]{biblatex}
\usepackage{amsmath}
\usepackage{amsfonts}
\usepackage{enumitem}
\usepackage{titlesec}
\usepackage{leftidx}
\usepackage{amsthm}
\usepackage{mathtools}
\usepackage{comment}
\usepackage{xcolor}
\allowdisplaybreaks
\titlelabel{\thetitle.\quad}
\usepackage[margin=1in]{geometry}

\newcommand{\opt}{\ensuremath{\mathsf{Opt}}}

\newcommand{\cost}{\ensuremath{\mathsf{Cost}}}
\newcommand{\ind}{\ensuremath{\mathsf{Ind}}}
\newcommand{\errf}{\ensuremath{\mathsf{Error}}}

\newcommand{\RR}{\mathbb{R}}

\newcommand{\half}{\frac{1}{2}}

\newcommand\abs[1]{\left\lvert#1\right\rvert}

\newcommand\parens[1]{\left(#1\right)}
\newcommand\brackets[1]{\left[#1\right]}
\newcommand*{\pr}[2][]{\text{Pr}\ifx\\\left[#1\right]\\\else_{#1}\fi \left[#2\right]}
\newcommand*{\EE}[2][]{\mathbb{E}\ifx\\\left[#1\right]\\\else_{#1}\fi \left[#2\right]}
\newcommand{\err}{\ensuremath{\mathsf{Err}}}
\newcommand{\nstop}{\ensuremath{n_{\mathsf{stop}}}}

\newcounter{theorem}
\newtheorem{defin}{Definition}[section]

\newtheorem{lemma}[defin]{Lemma}
\newtheorem{theorem}[defin]{Theorem}
\newtheorem{corollary}[defin]{Corollary}
\newtheorem{claim}[defin]{Claim}
\newtheorem{remark}[defin]{Remark}
\newtheorem{prop}[defin]{Proposition}
\newtheorem{fact}[defin]{Fact}
\newtheorem{proposition}[defin]{Proposition}

\begin{document}

\begin{titlepage}
\title{Binary Scoring Rules that Incentivize Precision}


\author{Eric Neyman, Georgy Noarov, S. Matthew Weinberg}


\maketitle

\begin{abstract}
All proper scoring rules incentivize an expert to predict \emph{accurately} (report their true estimate), but not all proper scoring rules equally incentivize \emph{precision}. Rather than treating the expert's belief as exogenously given, we consider a model where a rational expert can endogenously refine their belief by repeatedly paying a fixed cost, and is incentivized to do so by a proper scoring rule. 

Specifically, our expert aims to predict the probability that a biased coin flipped tomorrow will land heads, and can flip the coin any number of times today at a cost of $c$ per flip. Our first main result defines an \emph{incentivization index} for proper scoring rules, and proves that this index measures the expected error of the expert's estimate (where the number of flips today is chosen adaptively to maximize the predictor's expected payoff). Our second main result finds the unique scoring rule which optimizes the incentivization index over all proper scoring rules. 

We also consider extensions to minimizing the $\ell^{th}$ moment of error, and again provide an incentivization index and optimal proper scoring rule. In some cases, the resulting scoring rule is differentiable, but not infinitely differentiable. In these cases, we further prove that the optimum can be uniformly approximated by polynomial scoring rules.

Finally, we compare common scoring rules via our measure, and include simulations confirming the relevance of our measure even in domains outside where it provably applies.
\end{abstract}

\end{titlepage}

\section{Introduction}\label{sec:intro}

In the context of decision theory, a \emph{scoring rule} rewards predictors for the accuracy of their predictions~\cite{Good, Brier, Savage}. In the context of a binary choice (e.g. ``Will it rain tomorrow?"), a scoring rule can be thought of as a function $f: (0, 1) \to \RR$, where if a predictor reports a probability $p$ of rain, then the predictor's reward is $f(p)$ if it rains and $f(1 - p)$ if it does not rain.\footnote{To be clear: if $f(x) = \ln(x)$ and a predictor predicts a probability of $0.7$ to it raining, then the predictor receives reward $\ln(0.7)$ if it rains and $ \ln(0.3)$ if it does not rain.} We consider settings in which there are two possible outcomes that are treated symmetrically (as this definition assumes), and henceforth refer to scoring rules in terms of this function $f$. Traditionally, scoring rules are concerned with incentivizing \emph{accurate} reports. For example, a scoring rule is called \emph{proper} if a predictor is always incentivized to tell the truth, in the sense that reporting the predictor's true belief strictly maximizes the predictor's expected reward. 

Of course, there is an extraordinary amount of flexibility in selecting a proper scoring rule. For example, if a continuously differentiable scoring rule $f: (0, 1) \to \RR$ satisfies $xf'(x) = (1 - x)f'(1 - x)$ and $f'(x) > 0$ for all $x$, then $f$ is proper. Any increasing $\mathcal{C}^1$ function on $[\half, 1)$ can therefore be extended to a $\mathcal{C}^1$ proper scoring rule on $(0, 1)$ (see Corollary~\ref{cor:strictlyproper}). Much prior work exists comparing proper scoring rules by various measures, e.g.~\cite{Winkler1996, GneitingR07,Dawid2014}, but there is little which formally analyzes the extent to which proper scoring rules incentivize \emph{precision} (see Section~\ref{sec:related} for a discussion of prior work). 

As a motivating example, consider the problem of guessing the probability that one of two competing advertisements will be clicked. With zero effort, a predictor could blindly guess that each is equally likely. But the predictor is not exogenously endowed with this belief, they can also endogenously exert costly effort to refine their prediction. For example, the predictor could sample which ad they would click themselves, or poll members of their household for additional samples. A more ambitious predictor could run a crowdsourcing experiment, paying users to see which link they would click. Any proper scoring rule will equally incentivize the predictor to accurately report their resulting belief, but not all scoring rules equally incentivize the costly gathering of information.

We propose a simple model to formally measure the extent to which a scoring rule incentivizes costly refinement of the predictor's beliefs. Specifically, we consider a two-sided coin that comes up heads with probability $p$, and $p$ is drawn uniformly from $(0,1)$ (we refer to $p$ as the bias of the coin). Tomorrow the coin will be flipped, and we ask the predictor to guess the probability that it lands heads. Today, the predictor can flip the coin (with bias $p$) any number of times, at cost $c$ per flip. While we choose this model for its mathematical simplicity, it captures examples like the previous paragraph surprisingly well: tomorrow, a user will be shown the two advertisements (clicking one). Today, the predictor can run a crowdsourcing experiment and pay any number of workers $c$ to choose between the two ads. This simple model also captures weather forecasting using ensemble methods surprisingly well, and we expand on this connection in Appendix~\ref{app:weather}.

With this model in mind, consider the following two extreme predictions: on one hand, the predictor could never flip the coin, and always output a guess of $1/2$. On the other, the predictor could flip the coin infinitely many times to learn $p$ exactly, and output a guess of $p$. Note that both predictions are accurate: the predictor is truthfully reporting their belief, and that belief is correct given the observed flips. However, the latter prediction is more precise. All proper scoring rules incentivize the predictor to accurately report their true prediction in both cases, but different scoring rules incentivize the predictor to flip the coin a different number of times. More specifically, every scoring rule induces a different optimization problem for the predictor, thereby leading them to produce predictions of different quality. In this model, the key question we answer is the following: \emph{which scoring rules best incentivize the predictor to produce a precise prediction?} 

\subsection{Our Results}
Our first main result is the existence of an \emph{incentivization index}. Specifically, if $\errf_c(f)$ denotes the expected error that a rational predictor makes when incentivized by scoring rule $f$ with cost $c$ per flip, we give a closed-form index $\ind(f)$ with the following remarkable property: for all respectful (see Definition~\ref{def:respectful}) proper scoring rules $f$ and $g$, the inequality $\ind(f) < \ind(g)$ implies the existence of a sufficiently small $c_0 > 0$ such that $\errf_c(f) < \errf_c(g)$ for all $c \leq c_0$ (Theorem~\ref{thm:global}). We formally introduce this index in Definition~\ref{def:index}, but remark here that it is not a priori clear that such an index should exist at all, let alone that it should have a closed form.\footnote{Indeed, a priori it is possible that $\errf_{0.1}(f) < \errf_{0.1}(g)$, but $\errf_{0.01}(f) > \errf_{0.01}(g)$, and $\errf_{0.001}(f) < \errf_{0.001}(g)$, but $\errf_{0.0001}(f) > \errf_{0.0001}(g)$, and so on. The existence of an incentivization index rules out this possibility.} 

With an index in hand, we can now pose a well-defined optimization problem: which proper scoring rule minimizes the incentivization index? Our second main result nails down this scoring rule precisely; we call it $g_{1,\opt}$ (see Theorem~\ref{thm:optimal}). 

We also extend our results to the $\ell^{th}$ moment for $\ell \geq 1$, where now $\errf_c^\ell(f)$ denotes the expected $\ell^{th}$ power of the error that a rational predictor makes when incentivized by $f$ with cost $c$ per flip, and again derive an incentivization index $\ind^\ell(f)$ and an optimal scoring rule $g_{\ell,\opt}$. 

Some optimal rules $g_{\ell,\opt}$ have a particularly nice closed form (for example, as $\ell \rightarrow \infty$, the optimal rule pointwise converges to a polynomial), but many do not. We also prove, using techniques similar to the Weierstrass approximation theorem~\cite{Weierstrass1885}, that each of these rules can be approximated by polynomial scoring rules whose incentivization indices approach the optimum. 

Finally, beyond characterizing the optimal rules, the incentivization indices themselves allow for comparison among popular scoring rules, such as logarithmic ($f_{\log}(x):= \ln(x)$), quadratic ($f_{\text{quad}}(x):=2x - (x^2 + (1-x)^2)$), and spherical ($f_{\text{sph}}(x):=x/\sqrt{x^2 + (1-x)^2}$). We plot the predictions made by our incentivization index (which provably binds only as $c \rightarrow 0$) for various values of $c$, and also confirm via simulation that the index has bite for reasonable choices of $c$.

\subsection{Related Work}\label{sec:related}
To the best of our knowledge,~\cite{Osband89} was the first to consider scoring rules as motivating the predictor to seek additional information about the distribution before reporting their belief. This direction is revisited in work of~\cite{Clemen}, and has gained more attention recently~\cite{Tsakas, RoughgardenS17, HartlineLSW20}. While these works (and ours) each study the same phenomenon, there is little technical overlap and the models are distinct: each explores a different aspect of this broad agenda. For example,~\cite{RoughgardenS17} considers the predictor's incentive to outperform competing predictors (but there is no costly effort --- the predictors' beliefs are still exogenous).~\cite{HartlineLSW20} (which is contemporaneous and independent of our work) is the most similar in motivation, but still has significant technical differences (beyond the two subsequent examples). On one hand, their model is more general than ours in that they consider multi-dimensional state spaces (rather than binary ones, in our model). On another hand, it is more restrictive in that they consider only two levels of effort (versus infinitely many, in our model). 

Our work also fits into the broad category of principal-agent problems. For example, works such as~\cite{CaiDP15,LiuC16,ChenILSZ18,ChenZ19} consider a learning principal who incentivizes agents to make costly effort and produce an accurate data point. Again, the models are fairly distinct, as these works focus on more sophisticated learning problems (e.g.~regression), whereas we perform a more comprehensive dive into the problem of simply eliciting the (incentivized-to-be-precise) belief.

In summary, there is a sparse, but growing, body of work addressing the study of incentivizing effort in forming predictions, rather than just accuracy in reporting them. The above-referenced works pose various models to tackle different aspects of this agenda. In comparison, our model is arguably the simplest, and we develop a deep understanding of optimal scoring rules in this setting.

\subsection{Summary and Roadmap}
Section~\ref{sec:prelim} lays out our model, and contains some basic facts to help build intuition for reasoning about the incentivization properties of scoring rules. Our main results are detailed in Sections~\ref{sec:index} through~\ref{sec:weierstrass}, along with intuition for our techniques. 
\begin{itemize}
\item Section~\ref{sec:index} defines the incentivization index, and provides a sufficient condition (Definition~\ref{def:respectful}) for the incentivization index to nail down the expected error of a rational predictor, up to $o(1)$. This is our first main result, which gives a framework to formally reason about scoring rules that incentivize precision.
\item Section~\ref{sec:optimal} finds the unique proper scoring rule which optimizes the incentivization index. This is our second main result, which finds novel scoring rules, and also sets a benchmark with which to evaluate commonly-studied scoring rules.
\item Section~\ref{sec:compare} studies the optimal scoring rules from Section~\ref{sec:optimal}, and compares their incentivization indices to those of some well-known scoring rules. Appendix~\ref{app:simulation} provides a few simulations confirming that $\ind$ seems to have predictive value for $c \gg 0$.
\item Section~\ref{sec:weierstrass} proves that there exist polynomial scoring rules with incentivization indices arbitrarily close to the optimum.
\item All sections additionally consider the expected $\ell^{th}$ power of the error for any $\ell \geq 1$.
\item Section~\ref{sec:conclusion} concludes.
\end{itemize}

\section{Model and Preliminaries}\label{sec:prelim}
\subsection{Scoring Rules and their Rewards} \label{sec:reward}
This paper considers predicting a binary outcome for tomorrow: heads or tails. The \emph{expert} or \emph{predictor} is asked to output a probability $p$ with which they believe the coin will land heads. Tomorrow, should the coin land heads, their reward is $f(p)$; should it not, their reward is $f(1-p)$ (note that the reward is symmetric: it is invariant under swapping the labels `heads' and `tails'). Throughout this paper, we consider a scoring rule to be defined by this function $f(\cdot)$. Observe that if the expert believes the true probability of heads to be $p$, and chooses to guess $x$, then the expected reward is $p\cdot f(x) + (1-p)\cdot f(1-x)$. 

\begin{defin}[Expected Reward] For scoring rule $f: (0,1)\rightarrow \RR$, denote by $r^f_p(x):= p\cdot f(x) + (1-p)\cdot f(1-x)$ the expected reward of an expert who predicts $x$ when their true belief is $p$. 

Let also $R^f(p):= r^f_p(p)$ be the expected reward of an expert who reports their true belief $p$. We may drop the superscript when the scoring rule $f$ is clear from context.
\end{defin} 

A scoring rule is (weakly) proper if it (weakly) incentivizes accurate reporting. In our notation:

\begin{defin}
A scoring rule $f: (0, 1) \to \RR$ is \emph{proper} (resp. \emph{weakly proper}) if for all $p \in (0, 1)$, the expected reward function $r^f_p(x)$ is strictly (resp. weakly) maximal at $x = p$ on $(0, 1)$.
\end{defin}

Note that the optimal scoring rules \emph{designed} in this paper are all (strictly) proper. However, we will show them to be optimal even with respect to the larger class of weakly proper scoring rules.

\subsection{Modeling the Expert's Behavior}\label{sec:incentives}
We model the expert as Bayesian. Specifically, the expert initially believes the coin bias is uniformly distributed in $(0,1)$. Today, the expert may flip the coin any number of times in order to gauge its true bias, and pays $c$ per flip. After having flipped the coin $n$ times, and seen $k$ heads, the expert believes the true bias is $\frac{k+1}{n+2}$ (Fact~\ref{fact:flipping}).\footnote{By this, we mean the expert believes the coin would land heads with probability $\frac{k+1}{n+2}$, if it were flipped again.} Once done flipping, the expert reports the coin bias. Tomorrow, the coin is flipped once, and the expert receives reward for the prediction based on the outcome via scoring rule $f$ (known to the expert in advance), as described in Section~\ref{sec:reward}. 

It remains to define when the expert should stop flipping. Below, an \emph{adaptive strategy} simply refers to a (possibly randomized) stopping rule for the expert, i.e. a rule that, given any number of past flips and the scoring rule $f$, tells the expert whether to stop or flip once again. The payoff of an adaptive strategy is simply the expert's expected reward for following that strategy, minus the expected number of coin flips.

\begin{defin}
A \emph{globally-adaptive expert} uses the payoff-maximizing adaptive strategy.
\end{defin}

Nailing down the expert's optimal behavior as a function of $c$ is quite unwieldy. Thus, we derive our characterizations up to $o(1)$ terms (as $c \rightarrow 0$). When $c$ is large, one may reasonably worry that these $o(1)$ terms render our theoretical results irrelevant. In Appendix~\ref{app:simulation} we simulate the expert's optimal behavior for large $c$, and confirm that our results hold qualitatively in this regime. 


Finally, we define a natural measure of precision for the expert's prediction.

\begin{defin} The \emph{expected error} associated with a scoring rule $f$ and cost $c$ is $\errf_c(f):= \EE{\abs{p - q}}$. The expectation is taken over $p$, drawn uniformly from $(0,1)$, and $q$, the prediction of a globally-adaptive expert after flipping the coin ($q$ is a random variable which depends on $f, p, c$).


We will also consider generalizations to other moments, and define $\errf^\ell_c(f):= \mathbb{E}[|p-q|^\ell]$.
\end{defin}

\subsection{Scoring Rule Preliminaries}
Our proofs will make use of fairly heavy single-variable analysis, and therefore will require making some assumptions on $f(\cdot)$: continuity, differentiability, but also more technical ones. We will clearly state them when necessary, and confirm that all scoring rules of interest satisfy them. For these preliminaries, we need only assume that $f$ is continuously differentiable so that everything which follows is well-defined. First, Lemma~\ref{lem:weaklyproper} provides an alternative characterization of proper (and weakly proper) scoring rules. The proof is in Appendix~\ref{app:prelim}.

\begin{lemma}\label{lem:weaklyproper}
A continuously differentiable scoring rule $f$ is weakly proper if and only if for all $p \in (0, 1)$, $pf'(p) = (1 - p)f'(1 - p)$ and $f'(p) \ge 0$. It is (strictly) proper if and only if additionally $f'(p) > 0$ almost everywhere\footnote{Almost everywhere on $(0,1)$ refers to the interval $(0,1)$ except a set of measure zero.} in $(0,1)$.
\end{lemma}

\begin{corollary}\label{cor:strictlyproper}
Let $f$ be strictly increasing almost everywhere (resp., nondecreasing everywhere) and continuously differentiable on $(0,\half]$. Then $f$ can be extended to a continuously differentiable proper (resp., weakly proper) scoring rule on $(0,1)$ by defining $f'(p) = \frac{1-p}{p}f'(1-p)$ for $p \in (\half,1)$.
\end{corollary}

Put another way: every continuously differentiable proper scoring rule can be defined by first providing a strictly increasing function on $(0,\half]$, and then extending it as in Corollary~\ref{cor:strictlyproper}. Remark~\ref{rem:example} provides a short example to help parse this extension.

\subsection{First Steps towards Understanding Incentivization}\label{sec:proper}
In this section, we state a few basic facts about the expert's expected reward, and how it changes with additional flips. We defer all proofs to Appendix~\ref{app:prelim}. Reading these proofs may help a reader gain technical intuition for the model. Our analysis will focus mostly on the reward function $R^f(\cdot)$ rather than $f(\cdot)$, so the following fact will be useful:

\begin{fact}\label{fact:basic} For a weakly proper scoring rule $f$, we have $(R^f)'(x) = f(x) - f(1-x)$ and $(R^f)''(x) = f'(x) + f'(1-x) = \frac{f'(x)}{1 - x} \geq 0$ on $(0,1)$.
\end{fact}

Lemma~\ref{lem:increward} observes how this expected reward evolves with an additional flip.

\begin{lemma}\label{lem:increward} If the expert has already flipped the coin $n$ times, seeing $k$ heads, then their expected increase in reward for exactly one additional flip is $\frac{k+1}{n+2}R^f(\frac{k+2}{n+3}) + \frac{n-k+1}{n+2}R^f(\frac{k+1}{n+3}) - R^f(\frac{k+1}{n+2})$.
\end{lemma}

Lemma~\ref{lem:increward} suggests that the function $R^f(x)$ should be convex: if it were not, that would leave open the possibility of the expert potentially \emph{losing} expected reward as a result of performing \emph{more} flips (meaning that the expert might get a smaller reward for a better estimate of the coin bias).

\begin{lemma}[\cite{mcc56}]\label{lem:convex} Let $f$ be any proper (resp., weakly proper) scoring rule. Then $R^f(x)$ is strictly convex (resp., weakly convex) almost everywhere  on $(0,1)$.
\end{lemma}

\begin{corollary}\label{cor:increward} Let $f$ be a proper (resp., weakly proper) scoring rule. Then the expert's increased expected reward from an additional flip is strictly positive (resp., weakly positive). 
\end{corollary}

Because we are interested in incentivizing the expert to take costly actions, the scale of a proper scoring rule will also be relevant. For example, if $f$ is proper, then so is $2f$, and $2f$ clearly does a better job of incentivizing the expert (Lemma~\ref{lem:increward}). As such, we will want to first \emph{normalize} any scoring rule under consideration to be on the same scale. A natural normalization is to consider two scoring rules to be on the same scale if expected payoff they provide to the expert is the same (where the expectation is taken over both the bias and the flips of the coin). 

\begin{defin}We define $\cost_c(f)$ to be the expected payoff to a globally-adaptive expert via scoring rule $f$ (when the bias is drawn uniformly from $(0,1)$, and the expert may pay $c$ per flip).
\end{defin}

Recall that the (expected) payoff of a perfect expert is $\int_0^1 R(x) dx$, since a perfect expert has expected payoff $R(x)$ if the coin has bias $x$, and the coin's bias is chosen uniformly from $[0, 1]$. For proper (but not necessarily weakly proper) scoring rules, we show that as $c \rightarrow 0$ the expected payoff of a globally-adaptive expert approaches the payoff of a perfect expert. (This is true no matter the coin's bias, though we only need this result in expectation over the bias.) Intuitively, this is because the number of flips approaches $\infty$ as $c \rightarrow 0$, so the expert is rewarded as if they are perfect. 

\begin{proposition}\label{prop:cost} Let $f$ be a proper scoring rule. Then $\lim_{c \rightarrow 0} \cost_c(f) = \int_0^1 R(x) dx$. That is, $\cost_c(f) = \int_0^1 R(x) dx \pm o(1)$.
\end{proposition}

Assuming that two scoring rules $f,g$ have $\cost_c(f) = \cost_c(g)$ addresses one potential scaling issue. But there is another issue as well: whenever $f$ is proper, the scoring rule $2f-1$ is also proper, and again clearly does a better job incentivizing the expert (again directly by Lemma~\ref{lem:increward}). As such, we will also normalize so that $R^f(x)\geq 0$ for all $x$: the expert's expected reward is always non-negative if they are perfect. We conclude this section with a formal statement of this normalization. Appendix~\ref{app:prelim} confirms the implications of the definition, and also contains a few lemmas stating equivalent conditions.

\begin{defin}\label{def:normalized} A scoring rule $f(\cdot)$ is \emph{normalized} if $\int_0^1 R^f(x)dx = 1$, and $f(1/2) = 0$. This implies that $\cost_c(f) = 1\pm o(1)$, and that a perfectly calibrated expert gets non-negative expected reward. It also implies that an expert who flips zero coins gets zero expected reward.
\end{defin}

\section{An Incentivization Index}\label{sec:index}
This section presents our first main contribution: an incentivization index which characterizes the expert's expected error. The main result of this section, Theorem~\ref{thm:global}, requires scoring rules to be analytically nice in a specific way. We term such scoring rules \emph{respectful}.

\begin{defin} \label{def:respectful}
A proper scoring rule $f$ with reward function $R:=R^f$ is \emph{respectful} if:
\begin{enumerate}[label=(\arabic*)]
    \item $R$ is strongly convex on $(0, 1)$. That is, $R''(x) \ge a$ on $(0, 1)$ for some $a > 0$.
    \item $R'''$ is Riemann integrable on any closed subinterval of $(0, 1)$.\footnote{Note this does not necessarily require $R'''$ be defined on the entire $(0,1)$, just that it is defined almost everywhere.}
    \item $ \exists t > \frac{1}{4}$, and $c_0 > 0$ s.th.~for all $c \in (0,c_0)$: $\abs{R'''(x)} \le \frac{1}{c^{0.16} \sqrt{x(1 - x)}}R''(x)$ on $[c^t, 1 - c^t]$.\footnote{Except in places where $R'''$ is undefined.} 
\end{enumerate}
\end{defin}


Recall that $R$ is strictly convex for any strictly proper scoring rule, so strong convexity is a minor condition. Likewise, the second condition is a minor ``niceness" assumption. We elaborate on the third condition in detail in Appendix~\ref{app:respect}, and confirm that frequently used proper scoring rules are indeed respectful. We briefly note here that intuitively, the third condition asserts that $R''$ does not change too quickly (except possibly near zero and one) for small enough coin-flipping costs $c$. The particular choice of $0.16$ is not special, and could be replaced with any constant $< 1/6$. 

\begin{defin}[Incentivization Index] \label{def:index} We define the \emph{incentivization index} of a scoring rule $f$:

\[\ind(f):= \int_0^1 \parens{\frac{x(1 - x)}{(R^f)''(x)}}^{1/4} dx. \quad \text{ More generally, for $\ell \geq 1$: } \ind^\ell(f):= \int_0^1 \parens{\frac{x(1 - x)}{(R^f)''(x)}}^{\ell/4} dx.\]
\end{defin}

\begin{theorem} \label{thm:global}
If $f$ is a respectful, continuously differentiable proper scoring rule, then:
\[\lim_{c \to 0} c^{-1/4} \cdot \errf_c(f) = \sqrt{2/\pi}\cdot 2^{1/4} \cdot \ind(f).\]

More generally, if $\mu_\ell := \frac{2^{\ell/2} \Gamma \parens{\frac{\ell + 1}{2}}}{\sqrt{\pi}}$ is the $\ell^{th}$ moment of the standard normal distribution, then:

\[\lim_{c \to 0} c^{-\ell/4}\cdot \errf^\ell_c(f) = \mu_\ell\cdot 2^{\ell/4} \cdot \ind^\ell(f).\]
\end{theorem}

Intuitively, the incentivization index captures the expert's error as $c \rightarrow 0$.\footnote{Proposition~\ref{prop:n_stop_bound} in Section~\ref{sec:three} gives intuition for why $\errf_c(f)$ is proportional to $\sqrt[4]{c}$.} More formally, for any two respectful proper scoring rules $f, g$, $\ind(f) < \ind(g)$ implies that there exists a sufficiently small $c_0 > 0$ such that $\errf_c(f) < \errf_c(g)$ for all $c \leq c_0$. As previously referenced, Theorem~\ref{thm:global} says nothing about how big or small this $c_0$ might be, although simulations in Appendix~\ref{app:simulation} confirm that it does not appear to be too small for typical scoring rules.

The rest of this section is organized as follows. Sections~\ref{sec:global} through~\ref{sec:five} outline our proof of Theorem~\ref{thm:global}. The key steps are given as precisely-stated technical lemmas with mathematical intuition alongside them, to illustrate where precision is needed for the proof to carry through. Complete proofs of these lemmas can be found in Appendix~\ref{app:index}. In Appendix~\ref{app:respect}, we confirm that natural scoring rules are respectful (which is mostly a matter of validating the third condition in Definition~\ref{def:respectful}).

\subsection{Proof Outline of Theorem~\ref{thm:global}}\label{sec:global}
We provide below an executive overview of our approach. The concrete steps are separated out as formally-stated technical lemmas in the following sections, with proofs deferred to Appendix~\ref{app:index}. Before beginning, we highlight the main challenge: to prove Theorem~\ref{thm:global}, we need to capture the \emph{precise asymptotics} of the expert's expected error. Upper bounds can be easily shown via concentration inequalities; however, traditional lower bounds via anti-concentration results would simply state that the expected error tends to $0$ as $c \rightarrow 0$ (which holds for every proper scoring rule, and doesn't distinguish among them). So not only are we looking for two-sided bounds on the error, but we need to gauge the precise \emph{rate} at which it approaches zero. Moreover, even obtaining the order of magnitude of the error as $c \to 0$, which turns out to be $c^{-\ell/4}$, still does not suffice: we need to compute the exact coefficient of $c^{-\ell/4}$. This difficulty motivates the need for the technical lemmas stated in this section to be very precise. Our outline is as follows:
\begin{itemize}
\item All of our analysis first considers a locally-adaptive expert, who flips the coin one additional time if and only if the expected increase in reward \emph{from that single flip} exceeds $c$.
\item Our first key step, Section~\ref{sec:one}, provides an asymptotic \emph{lower bound} on the number of times an expert flips the coin, for all respectful $f$. 
\item Our second key step, Section~\ref{sec:two}, provides a coupling of the expert's flips across all possible true biases $p$. This helps prove uniform convergence bounds over all $p$ for the expert's error: we can now define an unlikely ``bad" event of overly-slow convergence without reference to $p$.
\item Our third key step, Section~\ref{sec:three}, provides tight bounds on the number of flips by a locally-adaptive expert, up to $(1\pm o(1))$ factors. Note that the first three steps have not referenced an error measure at all, and only discuss the expert's behavior.
\item Our fourth key step, Section~\ref{sec:four}, shows how to translate the bounds in Section~\ref{sec:three} to tight bounds on the error of a locally-adaptive expert, again up to $(1\pm o(1))$ factors.
\item Finally our last step, Section~\ref{sec:five}, shows that the globally-adaptive expert behaves nearly-identically to the locally-adaptive expert, up to an additional $o(1)$ factor of flips.
\end{itemize}

We now proceed to formally state the main steps along this outline, recalling that the first several steps consider a locally-adaptive expert, whose definition is restated formally below:

\begin{defin}[Locally-Adaptive Expert]
The \emph{locally-adaptive expert} flips one more time if and only if making a \emph{single} additional coin flip (and then stopping) increases their expected payoff.
\end{defin}

\subsection{Step One: Lower Bounding Expert's Number of Flips}\label{sec:one}
We begin by tying the expert's expected marginal reward from one additional flip to $R''$. Below, $Q(n)$ denotes the random variable which is the expert's belief after $n$ flips. The important takeaway from Claim~\ref{claim:rdoubleprime} is that for fixed $n$, the expert's expected belief as a function of $Q(n)$ changes (roughly) as $Q(n)\cdot (1-Q(n)) \cdot R''(Q(n))$ --- this takeaway will appear in later sections. 

\begin{claim} \label{claim:rdoubleprime}
Let $\Delta_{n + 1}(q):=\mathbb{E}[R(Q(n+1))|Q(n) = q] - R(q)$ be the expected increase in the expert's reward (not counting the paid cost $c$) from the $(n+1)^{th}$ flip of the coin, given current belief $Q(n)=q$. Then there exist $c_1,c_2 \in [q - 1/n,q + 1/n]$ such that:
\[\Delta_{n + 1} = \frac{q\cdot (1 - q)}{2(n + 3)^2}(q\cdot R''(c_1) + (1 - q)\cdot R''(c_2))\]
\end{claim}

Recalling that the locally-adaptive expert decides to flip the coin for the $(n+1)^{th}$ time if and only if $\Delta_{n + 1} \ge c$, and assuming that $R''$ is bounded away from zero (Condition 1 in Definition~\ref{def:respectful}), we arrive at a simple lower bound on the number of coin flips.

\begin{claim} \label{claim:one_third_bound}
For all $f$ such that $(R^f)''$ is bounded away from zero, there exists $\alpha, c_0$ such that the expert is guaranteed to flip the coin at least $\frac{1}{\alpha c^{1/3}}$ times for all $c \leq c_0$ (no matter the true bias).
\end{claim}

Using basic concentration inequalities, Claim~\ref{claim:one_third_bound} immediately implies an asymptotic \emph{upper bound} on the expert's error. Recall, however, that we need a two-sided bound, and moreover that we need precise asymptotics of the error. Still, Claim~\ref{claim:one_third_bound} is the first step towards this.

\subsection{Step Two: Ruling Out Irregular Coin-Flipping Trajectories}\label{sec:two}
The expert's coin-flipping behavior depends on $Q(n)$, which depends on the fraction of realized coin flips which are heads, which itself depend on the coin's true bias $p$. Note, of course, that $Q(n)\rightarrow p$ as $n \rightarrow \infty$. If instead we had that $Q(n) = p$ \emph{exactly}, we could leverage Claim~\ref{claim:rdoubleprime} to better understand the number of flips as a function of $p$. Unfortunately, $Q(n)$ will not equal $p$ exactly, and it is even possible to have $Q(n)$ far from $p$, albeit with low probability. 

The challenge, then, is then how to handle these low-probability events, and importantly how to do so \emph{uniformly over $p$}. To this end, we consider the following coupling of coin-flipping processes over all possible biases. Specifically, rather than first drawing bias $p$ and then flipping coins with bias $p$, we use the following identically distributed procedure:
\begin{enumerate}[label=(\arabic*)]
\item Generate an infinite sequence $r_1, r_2, \dots$ of uniformly random numbers in $[0, 1]$.
\item Choose $p$ uniformly at random from $[0, 1]$. \label{step:choose_p}
\item For each $n$, coin $n$ comes up heads if and only if $r_n \le p$.
\end{enumerate}
Under this sampling procedure, $Q_p(n) := \frac{h_p(n) + 1}{n + 2}$ is the expert's estimate after flipping $n$ coins, where $h_p(n)$ is the number of heads in the first $n$ flips, if $p$ is the value chosen in step~\ref{step:choose_p}. 

With this procedure, we can now define a single bad event \emph{uniformly over all $p$}. Intuitively, $\Omega_N$ holds when, no matter what $p$ is chosen in step~\ref{step:choose_p}, the expert's Bayesian estimate of $p$ never strays too far from $p$ after $N$ flips. More formally, the complement of $\Omega_N$ is our single bad event:

\[\overline{\Omega_N} := \bigcup_{n = N}^\infty \bigcup_{j = 1}^{n - 1} \left\{\abs{Q_{j/n}(n) - \frac{j}{n}} > \frac{\sqrt{j(n - j)}}{2n^{1.49}} \right\}.\]
The expression on the right-hand side of the inequality can be rewritten as 
$\sqrt{\frac{\frac{j}{n} \parens{1 - \frac{j}{n}}}{n}} \cdot \frac{n^{.01}}{2}$,
where the radical term gives the order of the expected difference between $Q_{j/n}(n)$ and $\frac{j}{n}$. So intuitively, $\Omega_N$ holds unless the actual difference between $Q_{j/n}(n)$ and $\frac{j}{n}$ far exceeds its expected value. 

We have defined $\Omega_N$ so that, on the one hand, our subsequent analysis becomes tractable when $\Omega_N$ holds, and on the other hand, $\Omega_N$ fails to hold with probability small enough that our asymptotic results are not affected. Below, Claim~\ref{claim:omega_p} gives the property we desire from $\Omega_N$, and Claim~\ref{claim:omega_unlikely} shows that $\overline{\Omega_N}$ is unlikely. The key takeaway from Claim~\ref{claim:omega_p} is that when $\Omega_N$ holds, the expert's prediction is close to $p$ \emph{for all $n \geq N$ and $p \in (0,1)$} and this closeness \emph{shrinks with $n$}.

\begin{claim} \label{claim:omega_p}
The exists a sufficiently large $N_0$ such that for all $N\geq N_0$: if $\Omega_N$ holds, then \[\abs{Q_p(n) - p} \le \frac{\sqrt{p(1 - p)}}{n^{.49}} \quad \text{ for all } n \ge N \text{ and } p \in [1/n,1-1/n].\]
\end{claim}

\begin{claim} \label{claim:omega_unlikely}
\[\pr{\overline{\Omega_N}} = O \parens{e^{-N^{.01}}}.\]
\end{claim}

While it is trivial to see that $Q_p(n)$ approaches $p$ as $n \rightarrow \infty$, we reiterate that Claims~\ref{claim:omega_p} and~\ref{claim:omega_unlikely} guarantee quantitatively that: (a) when $\Omega_N$ holds, $|Q_p(n)-p|$ \emph{shrinks with $n$}, (b) the probability that $\Omega_N$ fails \emph{shrinks exponentially fast in $N$}, and (c) both previous bounds are \emph{uniform over $p$}. 

\subsection{Step Three: Tightly Bounding Expert's Number of Flips}\label{sec:three}

We now nail down the precise asymptotics of the number of the expert's flips as a function of the true bias $p$. This becomes significantly more tractable after assuming $\Omega_N$ holds. Below, the random variable $\nstop$ denotes the number of flips that a locally-adaptive expert chooses to make.

\begin{proposition} \label{prop:n_stop_bound}
Assume that $\Omega_N$ holds for some $N$, and let $t$ be as in Definition~\ref{def:respectful}. There exists a constant $\gamma$ and cost $c_0 > 0$ such that for all $c \leq c_0$ and all $p \in [2c^t, 1 - 2c^t]$, we have
\[\sqrt{\frac{p(1 - p)R''(p)}{2c}(1 - \gamma c^{1/300})} \le \nstop \le \sqrt{\frac{p(1 - p)R''(p)}{2c}(1 + \gamma c^{1/300})}.\]
\end{proposition}
Proposition~\ref{prop:n_stop_bound} has two key aspects. First, the upper and lower bounds on $\nstop$ match up to a $1\pm o(1)$ factor. Second, the $o(1)$ term is independent of $p$. To get intuition for why $\nstop \approx \sqrt{\frac{p(1-p)R''(p)}{2c}}$, recall that Claim~\ref{claim:rdoubleprime} shows after $n$ flips, the expected marginal gain is $\Delta_{n + 1} \approx \frac{p(1 - p)}{2n^2} R''(p)$. This quantity first falls below $c$, the cost per flip, after $n = \sqrt{\frac{p(1 - p)R''(p)}{2c}}$ flips.

\subsection{Step Four: Translating Number-of-Flips Bounds to Error Bounds}\label{sec:four}
Having pinned down $\nstop$ quite precisely, we will now obtain a tight bound on the error of the locally-adaptive expert's reported prediction. By contrast, the previous three steps performed an analysis of the locally-adaptive expert's coin-flipping behavior, which does not depend on the choice of error metric. Lemma~\ref{lem:close} below is a formal statement of the main step of this process, which nails down the asymptotics of the error conditioned on $\Omega_N$. Below, $\err_c(p)$ denotes a random variable equal to the locally-adaptive expert's error when the cost is $c$ and the true bias is $p$ (and the scoring rule $f$ is implicit). 

\begin{lemma} \label{lem:close}
Let $\ell\ge 1$ and $\mu_\ell := \frac{2^{\ell/2} \Gamma \parens{\frac{\ell + 1}{2}}}{\sqrt{\pi}}$ be the $\ell^{th}$ moment of a standard Gaussian. Let $N = \frac{1}{\alpha c^{1/3}}$ (so $N$ is implicitly a function of $c$). For all $p \in [2c^t, 1 - 2c^t]$ we have
\[(1 - o(1))\cdot  \mu_\ell \cdot \parens{\frac{2p(1 - p)}{R''(p)}}^{\ell/4} \le c^{-\ell/4} \cdot \EE{(\err_c(p))^\ell \mid \Omega_N} \le (1 + o(1)) \cdot\mu_\ell \cdot \parens{\frac{2p(1 - p)}{R''(p)}}^{\ell/4}\]
where the $o(1)$ term is a function of $c$ (but not $p$) that approaches zero as $c$ approaches zero.
\end{lemma}

Lemma~\ref{lem:close} is the key, but far from only, step in translating Proposition~\ref{prop:n_stop_bound} to tight bounds on the locally-adaptive expert's error. Intuitively, it states that the value of the expert's error will be, up to a $1\pm o(1)$ factor, consistent with what one would expect from using a quantitative central limit theorem in conjunction with the bound on \nstop\ from Proposition~\ref{prop:n_stop_bound}.

\subsection{Step Five: From Locally-Adaptive to Globally-Adaptive Behavior}\label{sec:five}
Finally, we extend our previous analysis from locally-adaptive to globally-adaptive experts. In particular, for a scoring rule that gives finite expected reward to a perfect expert, we prove that the globally-adaptive expert does not flip significantly more than a locally-adaptive expert would, and therefore their achieved errors are equal up to a $1\pm o(1)$ factor. Below, the random variable $n_g$ denotes the number of flips by the globally-adaptive expert.

\begin{lemma} \label{lem:global_helper}
Assume $f$ is respectful and normalizable (i.e. $\int_0^1 R(x) dx < \infty$). Let $\gamma$ be as in Proposition~\ref{prop:n_stop_bound}. There exists a $c_0 > 0$, such that for all $c \leq c_0$: If $\Omega_{\nstop}$ holds and $4c^t \le Q(\nstop) \le 1 - 4c^t$, then \[\nstop \leq n_g \le (1 + 6\gamma c^{1/300}) \nstop.\]
\end{lemma}

Lemma~\ref{lem:global_helper} is the key step in this portion of the analysis. The remaining work is to bound the impact of negligible events (such as $\Omega_{\nstop}$ failing, or $ Q(\nstop)$ being extremely close to $0$ or $1$) on our analysis. This completes our outline of the proof of Theorem~\ref{thm:global} (and we refer the reader back to Section~\ref{sec:global} for a reminder of this outline).
\section{Finding Optimal Scoring Rules} \label{sec:optimal}
Now that we have shown that the incentivization index characterizes how well any respectful scoring rule incentivizes a globally-adaptive expert to minimize error, we have a well-defined optimization problem: \emph{which normalized proper scoring rule has the lowest incentivization index} (and therefore minimizes the expert's expected error)? Recall the following necessary and sufficient set of conditions for a continuously differentiable and normalized scoring rule $g(\cdot)$ to be weakly proper:\footnote{Including weakly proper scoring rules in our optimization domain makes the analysis simpler. The optimal scoring rules are in fact strictly proper.}
\begin{itemize}
\item (Lemma~\ref{lem:weaklyproper}) For all $x \in (0, 1)$, $xg'(x) = (1 - x)g'(1 - x)$ and $g'(x) \ge 0$.
\item (Definition~\ref{def:normalized}, Corollary~\ref{cor:normalized}) $g \parens{\half} = 0$.
\item (Definition~\ref{def:normalized}, Corollary~\ref{cor:normalized}) $\int_\half^1 (1 - x)g'(x) dx = 1$.
\end{itemize}

So our goal is just to find the scoring rule which satisfies these constraints and minimizes the incentivization index:
\[\ind^\ell(g) = \int_0^1 \parens{\frac{x(1 - x)}{R''(x)}}^{\ell/4} dx = \int_0^1 \parens{\frac{x(1 - x)^2}{g'(x)}}^{\ell/4} dx.\]

The main result of this section is the following theorem:

\begin{theorem}\label{thm:optimal}
The unique continuously differentiable normalized proper scoring rule which minimizes $\ind^\ell(g)$ is:
\[g_{\ell,\opt}(x) = \begin{cases}\kappa_\ell \int_\half^x (t^{\ell - 8} (1 - t)^{2\ell + 4})^{1/(\ell + 4)} dt & x \le \half \\
\kappa_\ell \int_\half^x (t^\ell (1 - t)^{2\ell - 4})^{1/(\ell + 4)} dt & x \ge \half.\end{cases}\]
\end{theorem}

While $g_{\ell,\opt}$ is certainly challenging to parse, importantly it is a closed form, and can thus be numerically evaluated (and, it is provably optimal). A complete proof of Theorem~\ref{thm:optimal} appears in Appendix~\ref{app:optimal}. Appendix~\ref{app:plots} contains several plots of these scoring rules, alongside traditional ones. Section~\ref{sec:compare} immediately below also gives further discussion of these rules.

\section{Comparing Scoring Rules}\label{sec:compare}
In this section we compare various scoring rules by their incentivization indices, for various values of $\ell$. Of particular interest are the values $\ell = 1$ (expected absolute error), $\ell = 2$ (expected squared error), and the limit as $\ell \to \infty$ (which penalizes bigger errors ``infinitely more'' than smaller ones, so this regime corresponds to minimizing the probability of being very far off).

\subsection{Optimal Scoring Rules for Particular Values of $\ell$} \label{sec:compare_opt}
We begin by noting some values of $\ell$ for which the function $g_{\ell, \opt}$ takes a nice closed form. $\ell = 1$ happens to not be one such value. For $\ell = 2,4,8$, the functions $g_{\ell,\opt}$ can be written in terms of elementary functions on the entire interval $(0,1)$. For $\ell = 2$, the closed form on $(1/2,1)$ is a polynomial, although its extension via Corollary~\ref{cor:strictlyproper} to $(0,1/2)$ is not. For $\ell = 8$, the closed form on both $(0,1/2)$ and $(1/2,1)$ is a polynomial, although they are different. Interestingly, as $\ell \rightarrow \infty$, the closed form converges pointwise to a single polynomial. Specifically, for these values of $\ell$:

\textbf{For $\mathbf{\ell = 2}$:} On $[\half, 1)$, we have
\[g_{2, \opt}(x) = \kappa_2 \int_\half^x t^{2/3} dt = \frac{3}{5} \kappa_2\parens{x^{5/3} - \parens{\half}^{5/3}}.\]

\textbf{For $\mathbf{\ell = 8}$:} On $(0, \half]$, we have
\[g_{8, \opt}(x) = \kappa_8 \int_\half^x (1 - t)^{5/3} dt = \frac{3}{8} \kappa_8 \parens{\parens{\half}^{8/3} - (1 - x)^{8/3}}\]
and on $[\half, 1)$, we have
\[g_{8, \opt}(x) = \kappa_8 \int_\half^x (t^{2/3} - t^{5/3}) dt = \kappa_8 \parens{\frac{3}{5}\parens{x^{5/3} - \parens{\half}^{5/3}} - \frac{3}{8} \parens{x^{8/3} - \parens{\half}^{8/3}}}.\]

Finally, \textbf{as $\mathbf{\ell \to \infty}$}: on the entire interval $(0, 1)$, $g_{\ell, \opt}$ pointwise converges to
\[\lim_{\ell \to \infty} \kappa_\ell \cdot \int_\half^x t(1 - t)^2 dt = \frac{320}{3} \parens{\frac{1}{4}x^4 - \frac{2}{3} x^3 + \frac{1}{2}x^2 - \frac{11}{192}} = \frac{5}{9}(48x^4 - 128x^3 + 96x^2 - 11).\]

We refer to this last rule as $g_{\infty, \opt}$. Intuitively, minimizing the expected value of error raised to a power that approaches infinity punishes any error infinitely more than an even slightly smaller error. Put otherwise, this metric judges a scoring rule by the maximum (over $p \in (0, 1)$) of the spread of the distribution of expert error. The scoring rule $g_{\infty, \opt}$ has a very special property, which is that the quantity $\frac{x(1 - x)}{R''(x)} = \frac{x(1 - x)^2}{g_{\infty, \opt}'(x)}$, which appears in the incentivization index, is a constant regardless of $x$. This means that, in the limit as $c \to \infty$, the distribution of the expert's error is the same regardless of $p$. It makes intuitive sense that making the spread of the distribution of expert error uniform over all $p$ also minimizes the maximum of these spreads, which explains why $g_{\infty, \opt}$ has this interesting property.

As some of these rules are not infinitely differentiable, a natural question to ask is: what infinitely differentiable normalized function minimizes $\ind^\ell$? While (as we have shown by virtue of $g_{\ell, \opt}$ being the unique minimizer) achieving an incentivization index equal to $\ind^\ell(g_{\ell, \opt})$ with an infinitely differentiable scoring rule is impossible, it turns out that it is possible to get arbitrarily close --- and in fact it is possible to get arbitrarily close with \emph{polynomial} scoring rules. The main idea of the proof is to use the Weierstrass approximation theorem to approximate $g_{\ell, \opt}$ with polynomials. See Section~\ref{sec:weierstrass} for a full proof.

\subsection{Comparison of Incentivization Indices of Scoring Rules} \label{sec:comparison_ind}
We compare commonly studied scoring rules such as quadratic, logarithmic, and spherical, and refer to their normalizations as $g_\text{quad},g_\text{log},g_\text{sph}$, respectively. Additionally we include for comparison the normalization $g_\text{hs}$ of the $hs$ scoring rule, defined as $hs(x) = -\sqrt{\frac{1 - x}{x}}$. This scoring rule was prominently used in \cite{bb20} to prove their minimax theorem for randomized algorithms. 

Figure~\ref{tab:first} states $\ind^\ell(g)$ for various scoring rules $g$ (the lower the better).
\begin{figure}
\begin{center}
\begin{tabular}{r||c|c|c}
$\ind^\ell(\cdot)$ & $\ell = 1$ & $\ell = 2$ & $\ell = 4$\\
\hline
$g_\text{log}$ & 0.260 & 0.0732 & 0.00644\\
$g_\text{quad}$ & 0.279 & 0.0802 & 0.00694\\
$g_\text{sph}$ & 0.296 & 0.0889 & 0.00819\\
$g_\text{hs}$ & 0.255 & 0.0723 & 0.00658\\
$g_{1,\opt}$ & 0.253 & 0.0728 & 0.00719\\
$g_{2,\opt}$ & 0.255 & 0.0718 & 0.00661\\
$g_{4,\opt}$ & 0.261 & 0.0732 & 0.00639\\
$g_{\infty,\opt}$ & 0.311 & 0.0968 & 0.00974
\end{tabular} 
\end{center}\caption{Rows correspond to scoring rules, and columns correspond to error measures.}\label{tab:first}
\end{figure}
Figure~\ref{tab:first} lets us compare the performance of various scoring rules by our metric for any particular value of $\ell$. However, as one can see, $\ind^\ell$ decreases as $\ell$ increases. This makes sense, since $\ind^\ell$ measures the expected $\ell$-th power of error. For this reason, if we wish to describe how a given scoring rule performs over a range of values of $\ell$, we need to normalize these values. We do so by taking the $\ell$-th root and dividing these values by the $\ell$-th root of the optimal (smallest) index (and take the inverse so that larger numbers are better). This gives us the following measure of scoring rule precision, which makes sense across different values of $\ell$:
\[\parens{\frac{\ind^\ell(g_{\ell, \opt})}{\ind^\ell(g)}}^{1/\ell}.\]
\begin{figure}
\begin{center}
\includegraphics[scale=.7]{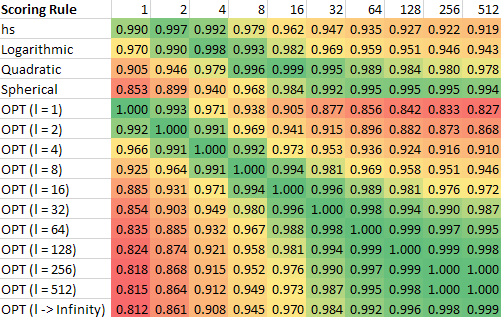}
\end{center}\caption{Rows correspond to different scoring rules $g$, and columns correspond to different measures of error $\ell$. The corresponding entry is $\parens{\frac{\ind^\ell(g_{\ell, \opt})}{\ind^\ell(g)}}^{1/\ell}$.}\label{fig:excel}
\end{figure} 

Figure~\ref{fig:excel}, which evaluates this expression for a selection of scoring rules and values of $\ell$, reveals some interesting patterns. Of the hs, logarithmic, quadratic, and spherical scoring rules, the hs scoring rule is the best one for the smallest values of $\ell$ and is in fact near-optimal for $\ell = 2$. The logarithmic rule is the best one for somewhat larger values of $\ell$ and is in near-optimal for $\ell \approx 4$. For larger values of $\ell$, the quadratic scoring rule is best, and is near-optimal for $\ell \approx 16$. For even larger values of $\ell$, the spherical scoring rule is the best of the four. This pattern suggests that for any given proper scoring rule there is a trade-off between incentivizing precision at low and at high values of $\ell$; it would be interesting to explore this further. 

Below is a continuous version of Figure~\ref{fig:excel}. The chart shows how the numbers above vary as $\ell$ ranges from $1$ to $200$.
\begin{center}
\includegraphics[scale=.7]{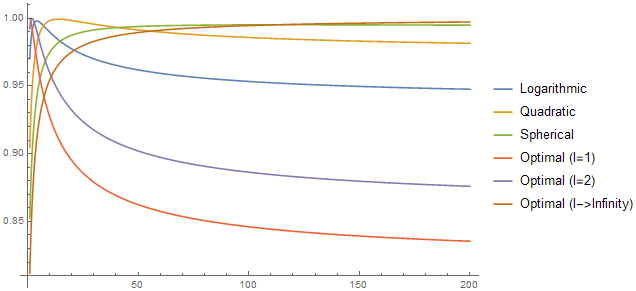}
\end{center}
And below is a zoomed-in version where $\ell$ ranges from $1$ to $10$.
\begin{center}
\includegraphics[scale=.7]{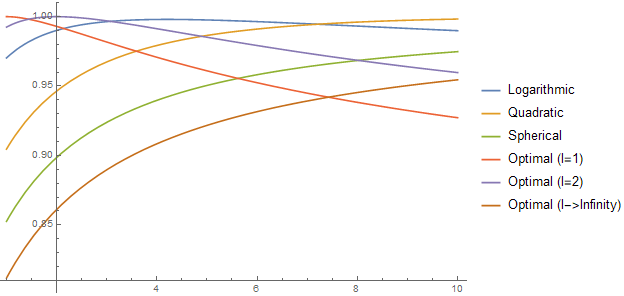}
\end{center}

\section{Almost-Optimal Incentivization Indices with Polynomial Respectful Scoring Rules} \label{sec:weierstrass}
The main result of this section is the following theorem, stating that polynomial,\footnote{To be clear, when we say a scoring rule $f(\cdot)$ is polynomial, we mean simply that $f(\cdot)$ is a polynomial function.} respectful proper scoring rules suffice to get arbitrarily close to the optimal incentivization index.

\begin{theorem} \label{thm:approx}
For $\ell \ge 1$ and $\varepsilon > 0$, there exists a respectful polynomial normalized proper scoring rule $f$ satisfying $\ind^\ell(f) \leq \ind^\ell(g_{\ell,\opt})+\varepsilon$.
\end{theorem}

The proof of Theorem~\ref{thm:approx} uses ideas from the Weierstrass approximation theorem. However, the Weierstrass approximation theorem gives a particular measure of ``distance'' between two functions, which does not translate to these functions having similar incentivization indices. So one challenge of the proof is ensuring convergence of a sequence of polynomials to $g_{\ell,\opt}$ \emph{in a measure related to $\ind^\ell$}. A second challenge is to ensure that all polynomials in this sequence are themselves proper, respectful scoring rules. Like previous technical sections, we include a few concrete lemmas to give a sense of our proof outline.

For example, one step in our proof is to characterize all \emph{analytic} proper scoring rules (that is, proper scoring rules that have a Taylor expansion which converges on their entire domain $(0,1)$). A necessary condition to be analytic is to be infinitely differentiable, which rules of the form $g_{\ell,\opt}$ are not, for any fixed $\ell$. We therefore seek to approximate such scoring rules with polynomial scoring rules (which are analytic), which are also respectful and proper.

\begin{theorem} \label{thm:analytic}
Let $f: (0, 1) \to \RR$ be analytic. Then $f$ is a proper scoring rule if and only if $f$ is nonconstant, $f'(x) \ge 0$ everywhere, and
\[f(x) = c_0 + \sum_{k > 0 \text{\emph{ odd}}} c_k(2k + 1 - 2kx)\parens{x - \half}^k\]
for some $c_0, c_1, c_3, c_5, \dots \in \RR$.
\end{theorem}

As an example to help parse Theorem~\ref{thm:analytic}, the quadratic scoring rule has $c_1 < 0$, and $c_i = 0$ for all other $i$. Using Theorem~\ref{thm:analytic}, we can conclude the following about $(R^f)''$ for any proper scoring rule $f$:

\begin{lemma} \label{lem:taylorphil}
Let $f: (0, 1) \to \RR$ be analytic. Then $f$ is a proper scoring rule if and only if $(R^f)''$ is not uniformly zero, nonnegative everywhere, and can be written as
\[(R^f)''(x) = \sum_{k \ge 0 \text{ even}} d_k \parens{x - \half}^k.\]
\end{lemma}

Lemma~\ref{lem:taylorphil} provides clean conditions on what functions $(R^f)''$ are safe to use in our sequence of approximations, and our proof follows by following a Weierstrass approximation-type argument while keeping track of these conditions. The rest of the details for the proof of Theorem~\ref{thm:approx} can be found in Appendix~\ref{app:weierstrass}.
\section{Conclusion}\label{sec:conclusion}

We propose a simple model, where an expert can expend costly effort to refine their prediction, and study the effectiveness of different scoring rules in incentivizing the expert to form a precise belief. Our first main result (Theorem~\ref{thm:global}) identifies the existence of a closed-form incentivization index: scoring rules with a lower index incentivize the expert to be more accurate. Our second main result (Theorem~\ref{thm:optimal}) identifies the unique optimal scoring rule with respect to this index. Section~\ref{sec:compare} then uses the incentivization index to compare common scoring rules (including our newly-found optimal ones), and Section~\ref{sec:weierstrass} shows that one can get arbitrarily close to the optimal incentivization index with polynomial scoring rules.

Our model is mathematically simple to describe, and yet it captures realistic settings surprisingly well (see Section~\ref{sec:intro} and Appendix~\ref{app:weather}). As such, there are many interesting directions for future work. For example:
\begin{itemize}
\item Our work considers a globally-adaptive expert, and establishes that they behave nearly identically to a locally-adaptive expert. What about a non-adaptive expert, who must decide a priori how many flips to make before seeing their results?
\item Our work considers a principal who wishes to minimize expected error. What if instead the principal wishes to optimize other objectives? In particular, are there objectives that are optimized by simpler rules (such as quadratic, logarithmic, etc.)? 
\item Our work considers optimal scoring rules for the incentivization index, and shows that polynomial scoring rules approach the optimum. Do \emph{exceptionally} simple scoring rules (such as quadratic, logarithmic, etc.) guarantee a good approximation to the optimal incentivization index for all $\ell$?
\end{itemize}

\printbibliography

\appendix
\section{Relationship of Model to Ensemble Weather Forecasts} \label{app:weather}
A major shift occurred in the field of weather forecasting around the turn of the 21st century. In the previous century, weather forecasting was viewed as inherently deterministic: a forecasting model would take as input some initial conditions and use differential equations to simulate future states of the atmosphere. A major complication of this approach, however, was that initial conditions are not perfectly known. While many weather stations observe conditions throughout the Earth, the chaotic nature of atmospheric phenomena meant that even small inaccuracies in initial conditions would produce substantial forecast inaccuracies even a few days into the future.

Starting in the early 1990s and continuing into the early 2000s, there was a paradigm shift away from deterministic forecasts and toward forecasts based on \emph{ensembles}. An ensemble is a run of a forecast model based on a perturbed set of initial conditions. Instead of making simulating the atmosphere starting from one ``best guess" set of initial conditions, ensemble-based forecasts would run some number --- generally between 5 and 100 --- ensembles, and use the results of these ensemble runs to generate a forecast~\cite{GR05}.

The initial conditions used in ensemble models are typically chosen by ``ensemble prediction systems," which attempt to sample the conditions from a probability distribution based on real-world uncertainty. The forecasts generated by ensemble models can treated as a sample from the probability distribution over the future weather. For instance, if 60\% of ensembles produce rain in New York seven days from now, a model might estimate the chance of rain in New York seven days from now at 60\%, perhaps slightly adjusted based on a prior inferred from historical climate data~\cite{GBR07}.

Each ensemble can be thought of as a coin flip whose cost is measured in time, energy, or computational resources. Each additional ensemble has a constant cost. The final forecast for a weather event is (roughly speaking) the fraction of ensembles that showed the event occurring. In this way, ensemble-based weather forecasting strongly parallels our coin flip based model of expert learning.
\section{Plots of Some Relevant Scoring Rules} \label{app:plots}
Below is a plot of $g_{\ell,\opt}$ for $\ell = 1, 2, 8, \infty$.
\begin{center}
\includegraphics[scale=0.85]{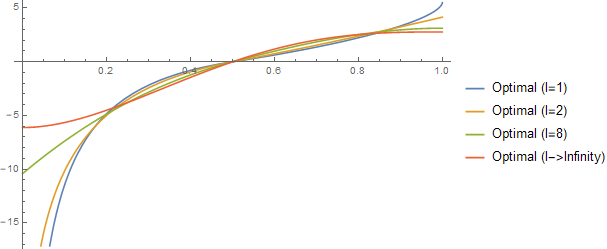}
\end{center}
As demonstrated by the plot, optimal scoring rules for larger values of $\ell$ are ``flatter," choosing to sacrifice rewarding precision near $0$ and $1$, in favor of rewarding precision closer to $\frac{1}{2}$. An expert rewarded by $g_{\infty, \opt}$ does not particularly care to distinguish between 98\% and 99\% probabilities, since the scoring rule is basically flat near the tails; this is not the case for $g_{1, \opt}$. Conversely, because $g_{\infty, \opt}$ is steeper than $g_{1, \opt}$ near $\frac{1}{2}$, an expert cares more about differentiating between a 50\% and a 51\% chance if rewarded with $g_{\infty, \opt}$ than with $g_{1, \opt}$.

Another, perhaps more enlightening way to view these scoring rules is through the quantity $\sqrt{\frac{x(1 - x)}{R''(x)}}$. Up to a constant factor depending on the cost $c$ of a flip, this is the variance of the normal distribution that approximates the distribution of the expert's response if the true bias of the coin is $x$ (for small $c$) --- or, put otherwise, the expected squared error. Below is a plot of this quantity for a variety of the scoring rules we have discussed.
\begin{center}
\includegraphics[scale=0.85]{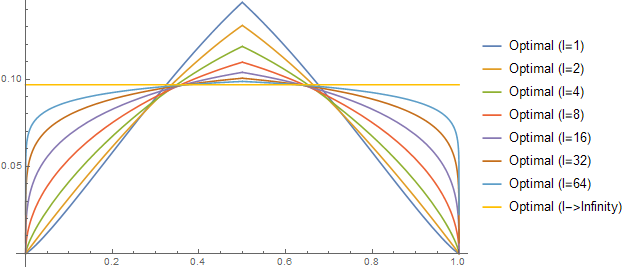}
\end{center}
This reinforces our previous point: optimal rules for small value of $\ell$ result in very small errors near $0$ and $1$, but relatively large errors in the middle. In Section~\ref{sec:compare_opt}, we discussed in brief why it makes sense that the value of $\frac{x(1 - x)}{R''(x)}$ is constant for the scoring rule $g_{\infty, \opt}$. This chart reinforces the point: since our normalization constraints force a trade-off between minimizing expert error for different values of the coin's bias $p$, the scoring rule whose error is independent of $p$ will have the minimum possible value of the maximum error over all $p$.

Finally, below is a plot like the previous one, but including the (normalized) logarithmic, quadratic, and spherical rules.
\begin{center}
\includegraphics[scale=0.85]{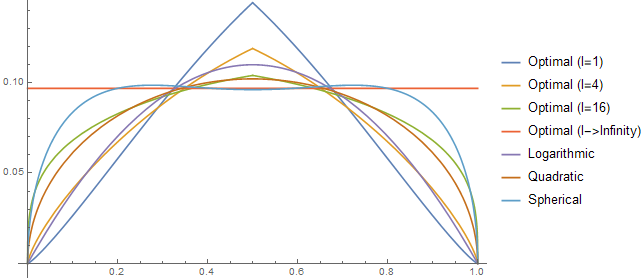}
\end{center}
In Section~\ref{sec:comparison_ind}, we noted that the logarithmic rule is near-optimal for $\ell \approx 4$, the quadratic rule for $\ell \approx 16$, and the spherical rule for even larger $\ell$. This plot helps provide some intuition: the log scoring rule is similar in shape to $g_{4, \opt}$ and similarly for the quadratic scoring rule and $g_{16, \opt}$, and for the spherical scoring rule and $g_{\infty, \opt}$.
\section{Omitted Proofs from Section~\ref{sec:prelim}}\label{app:prelim}

\subsection{Some Mathematical Preliminaries}\label{sec:math}
We will use the following mathematical facts throughout our proofs, and include them here for reference. 

\begin{fact}\label{fact:beta} $\int_0^1 x^k (1-x)^{n-k}dx = \frac{1}{n+1}/\binom{n}{k}$.
\end{fact}
\begin{proof}
Consider computing the probability of the following event in two different ways: there are $n+1$ values drawn independently and uniformly at random from $[0,1]$, $x_1,\ldots, x_{n+1}$. The event occurs if and only if the first $k$ elements are all smaller than the $(k+1)^{st}$, which is smaller than all of the last $n-k$ elements. 

One way to compute this probability is to first sample $x_{k+1}$ uniformly from $[0,1]$, and then compute the probability that each of the first $k$ elements are all smaller, and each of the last $n-k$ are all larger. This probability is exactly $\int_0^1 x^k (1-x)^{n-k}dx$. 

Another way is to first draw the $n+1$ values, and then sample a random permutation to map them to $x_1,\ldots, x_{n+1}$. Then for any $n+1$ values, the event occurs if and only if the $(k+1)^{st}$ smallest element is mapped to $x_{k+1}$, and then that the smallest $k$ values are mapped to $x_1,\ldots, x_k$. This happens with probability $\frac{1}{n+1}/\binom{n}{k}$. 
\end{proof}

\begin{fact}\label{fact:flipping} Say that the expert has flipped the coin $n$ times, and $k$ of them were heads. Then the expert believes that the probability of heads is $\frac{k+1}{n+2}$.
\end{fact}

\begin{proof}
This follows from an application of Bayes rule. The probability of seeing $k$ of $n$ heads, conditioned on the true bias being $p$ is $\binom{n}{k} p^k (1-p)^{n-k}$. Therefore, the probability of seeing $k$ of $n$ heads (unconditioned) is $\binom{n}{k} \int_0^1 p^k (1-p)^{n-k}dp = \frac{1}{n+1}$ (by Fact~\ref{fact:beta}). Therefore, the density that the true bias is $p$, conditioned on seeing $k$ of $n$ heads is $(n+1)\binom{n}{k}p^k (1-p)^{n-k}$, and the probability of seeing heads on the next flip is:

$$(n+1)\binom{n}{k} \int_0^1 p^{k+1}(1-p)^{n-k}dp =\frac{ (n+1)\binom{n}{k}}{(n+2)\binom{n+1}{k+1}} = \frac{k+1}{n+2}.$$
\end{proof}

\subsection{Omitted Proofs}

\begin{proof}[Proof of Lemma~\ref{lem:weaklyproper}]
We first prove that if $f$ is weakly proper then it satisfies the two stated constraints. Suppose that $f$ is weakly proper. It is clear that $f$ satisfies the first equation: for all $p$, in order for $r_p(x) = pf(x) + (1 - p)f(1 - x)$ to have a maximum at $x = p$, its derivative $pf'(x) - (1 - p)f'(1 - x)$ must be $0$ at $x = p$. So we first conclude that we must have $xf'(x) = (1-x)f'(1-x)$ for all $x \in (0,1)$. Next, observe that
\begin{equation} \label{rpxeq}
r_p'(x) = p f'(x) - (1-p)f'(1-x) = pf'(x) - (1 - p) \frac{xf'(x)}{1 - x} = f'(x) \parens{p - (1 - p) \frac{x}{1 - x}}.
\end{equation}
Suppose for contradiction that for some $p \in (0, 1)$, we have $f'(p) < 0$. Since $f'$ is continuous, $f'(x) < 0$ on some open interval containing $p$. On that open interval, then, the sign of $r_p'(x)$ is the opposite of the sign of $p - (1 - p) \frac{x}{1 - x}$ --- that is, negative when $x < p$ and positive when $x > p$. But then $r_p(x)$ is strictly \emph{minimized}, rather than maximized, at $x = p$ on this interval, contradicting that $f$ is weakly proper.

To prove the stronger claim when $f$ is proper, assume for contradiction that $f'(p)$ is not strictly positive almost everywhere. Then because $f'(\cdot)$ is continuous, there is an interval of non-zero length in which $f'(x) = 0$ on the entire interval. Let $p$ lie on the interior of this interval. Equation~\eqref{rpxeq} then establishes that $r_p'(x)$ is $0$ in an interval around $x = p$, meaning that $p$ is not the unique maximizer for $r_p(\cdot)$, contradicting that $f$ is proper.

Conversely, suppose that $f$ satisfies the two stated constraints. We show that $f$ is weakly proper by showing a stronger statement: that for all $p$, $r_p(x)$ weakly increases on $(0, p]$ and weakly decreases on $[p, 1)$. By the first constraint, (\ref{rpxeq}) holds. By the second constraint, for all $x$, $r_p'(x)$ is either $0$ or has the sign of $p - (1 - p)\frac{x}{1 - x}$, i.e. positive if $x < p$ and negative if $x > p$. This means that $r_p(x)$ is weakly increasing on $(0, p]$ and weakly decreasing on $[p, 1)$, and so attains a weak global maximum at $x = p$, as desired.

To prove the stronger claim when $f'(x) > 0$ almost everywhere, we show that $r_p(x)$ strictly increases almost everywhere on $(0, p]$ and strictly decreases almost everywhere on $[p, 1)$. Again, (\ref{rpxeq}) holds, so by the second constraint we have that $r_p'(x)$ has the sign of $p - (1 - p)\frac{x}{1 - x}$ almost everywhere, i.e. positive if $x < p$ and negative if $x > p$. Thus, $r_p(x)$ is strictly increasing almost everywhere on $(0, p]$ and strictly decreasing almost everywhere on $[p, 1)$, and so attains a strict global maximum at $x = p$, as desired.
\end{proof}

\begin{proof}[Proof of Corollary~\ref{cor:strictlyproper}]
As $\frac{1 - p}{p}$ is strictly positive on $(\half, 1)$, and $f'(1-p)$ is strictly (resp., weakly) positive almost everywhere on $(\half,1)$, we immediately conclude that $f'(p)$ is also strictly (resp., weakly) positive almost everywhere on $(\half, 1)$. Therefore, $f$ is proper (resp., weakly proper) by Lemma~\ref{lem:weaklyproper}.
\end{proof}

This extension will be relevant when we design optimal scoring rules, so we provide a quick example to help parse it. 

\begin{remark}\label{rem:example} Consider the function $f(x)=x$, which is strictly increasing on $(0,\half]$. Defining $f'(x) = \frac{1 - x}{x}\cdot 1 = \frac{1}{x} - 1$ for $x \in [\half, 1)$ results in $f(x) = \ln x - x + 1 + \ln 2$ (where $1 + \ln 2$ is the necessary constant to make $f$ continuous at $x = \half$). Clearly $f'(x) > 0$ on $(0, 1)$ (as promised by Corollary~\ref{cor:strictlyproper}), so we have just constructed a proper scoring rule:
\[f(x) = \begin{cases}x & \text{for } x \le \half \\ \ln x - x + 1 + \ln 2 & \text{for } x \ge \half\end{cases}.\]
\end{remark}

\begin{proof}[Proof of Fact~\ref{fact:basic}]
$R^f(x):=xf(x) + (1-x)f(1-x)$, so $(R^f)'(x) = xf'(x) +f(x) - f(1-x) - (1-x)f'(1-x) = f(x) - f(1 - x)$ by Lemma~\ref{lem:weaklyproper}. Taking the derivative again, we have $(R^f)''(x) = f'(x) + f'(1-x)$. Again by Lemma~\ref{lem:weaklyproper}, it follows that $(R^f)''(x) = \frac{f'(x)}{1 - x}$. Since $f'(x) \geq 0$ everywhere, we conclude the final inequality.
\end{proof}

\begin{proof}[Proof of Lemma~\ref{lem:increward}]
This is a direct application of Fact~\ref{fact:flipping}, which states that if the expert has flipped $h$ heads of $n$ coin, their guess at the coin's bias is $\frac{h + 1}{n + 2}$. Currently, the expert believes the probability of heads to be $\frac{k+1}{n+2}$. So their expected reward if they stop flipping now is exactly $R(\frac{k+1}{n+2})$. If they flip once more and stop, then with probability $\frac{k+1}{n+2}$ they will get a heads, updating their belief to $\frac{k+2}{n+3}$, and yielding expected reward $R(\frac{k+2}{n+3})$. With probability $\frac{n-k+1}{n+2}$ they will get a tails, updating their belief to $\frac{k+1}{n+3}$ and yielding expected reward $R(\frac{k+1}{n+3})$.
\end{proof}

\begin{proof}[Proof of Lemma~\ref{lem:convex}]
$R(x):= x\cdot f(x) + (1-x)\cdot f(1-x)$, so we get that $R'(x)= f(x) -f(1-x)+ xf'(x) -(1-x)f'(1-x) = f(x) - f(1-x)$ (by Lemma~\ref{lem:weaklyproper}). Therefore:

$$R''(x) = f'(x) +f'(1-x) \geq 0.$$

The final inequality also follows from Lemma~\ref{lem:weaklyproper}. If we further assume that $f$ is (strictly) proper, then we additionally get that $R''(x) > 0$ almost everywhere by Lemma~\ref{lem:weaklyproper}, as desired.
\end{proof}

\begin{proof}[Proof of Corollary~\ref{cor:increward}]
Observe that $\frac{k+1}{n+2}\cdot \frac{k+2}{n+3} + \frac{n-k+1}{n+2}\cdot \frac{k+1}{n+3} = \frac{(k+1)(k+2) + (n-k+1)(k+1)}{(n+2)(n+3)} = \frac{k+1}{n+2}$. Therefore, by Lemma~\ref{lem:convex}, $\frac{k+1}{n+2}R(\frac{k+2}{n+3}) + \frac{n-k+1}{n+2}R(\frac{k+1}{n+3}) -R(\frac{k+1}{n+2}) > 0$. 
\end{proof}

\begin{proof}[Proof of Proposition~\ref{prop:cost}]
First, we wish to argue that as $c \rightarrow 0$, \emph{no matter the true bias}, the number of flips any expert will choose to make approaches $\infty$. To see this, observe that after $n$ flips, the expert's current belief will always be an integer multiple of $\frac{1}{m+1}$. So if $c(m):= \min_{k<n \leq m+1, \text{ and } k,n \in \mathbb{N}}\{\frac{k+1}{n+2}R(\frac{k+2}{n+3}) + \frac{n-k+1}{n+2}R(\frac{k+1}{n+3}) - R(\frac{k+1}{n+2})\}$, then the expert will always flip the coin at least $n$ times as long as $c \leq c(m)$. Observe also that the minimum is taken over finitely many terms, all of which are strictly positive, so $c(m)$ is strictly positive. Therefore, for all $n$, there is a sufficiently small $c(m) > 0$ such that whenever the cost is at most $c(m)$, the expert flips at least $n$ times no matter the true bias. Note that while these calculations are done for an adaptive expert, they hold for a non-adaptive expert as well because the non-adaptive expert would want to flip at least $n$ coins \emph{no matter the outcomes}. 

Now, let's consider the expected reward for an expert who makes exactly $n$ flips no matter what. For all $k$, such an expert sees $k$ total heads with probability $1/(m+1)$. And conditioned on seeing $k$ heads, the expert's expert reward tomorrow is $R(\frac{k+1}{m+2})$. Therefore, we can conclude:

\begin{lemma}\label{lem:nonadapt} 
For all $n$, the expected reward after $n$ coin flips is $\sum_{k=0}^n R(\frac{k+1}{n+2})/(n+1)$.
\end{lemma}

Now, we want to understand the limt of this sum as $n \rightarrow \infty$. Observe that for each $n$, the sum is a Riemann sum for the function $R(x)$ on $[0,1]$ (i.e. each $\frac{k+1}{m+2}$ lies inside the interval $[\frac{k}{m+1}, \frac{k+1}{m+1}]$). Therefore, the limit as $n \rightarrow \infty$ is just the Riemann integral, and we get:

$$\lim_{m \rightarrow \infty} \sum_{k=0}^m R(\frac{k+1}{m+2})/(m+1) = \int_0^1 R(x)dx.$$

So now we can conclude that the non-adaptive expert gets expected payoff $\int_0^1 R(x)dx + o(1)$. As $c \rightarrow 0$, the number of flips $n \rightarrow \infty$, and the expected payoff as $n \rightarrow \infty$ approaches $\int_0^1 R(x) dx$. For the adaptive experts, observe that by Lemmas~\ref{lem:increward} and~\ref{lem:convex} that as long as they flip the coin at least $n$ times with probability $1$, their expected reward is at least as large as if they flipped it exactly $n$ times with probability $1$. As their expected reward can certainly not exceed $\int_0^1 R(x) dx$ (as this is the reward of a perfect expert who knows exactly the bias), their expected reward must also approach $\int_0^1 R(x) dx$ as $n \rightarrow \infty$ (and therefore as $c \rightarrow \infty$ as well).
\end{proof}

\begin{lemma}\label{lem:normalized} For a continuously differentiable (weakly) proper scoring rule $f$, $R^f(x) \geq 0$ for all $x$ if and only if $R^f(1/2) = f(1/2) \geq 0$.
\end{lemma}

\begin{proof}[Proof of Lemma~\ref{lem:normalized}]
Recall that $(R^f)'(x) = f(x) - f(1-x)$ (Fact~\ref{fact:basic}). Therefore, $(R^f)'(1/2) = 0$. Also, by Lemma~\ref{lem:convex}, $R^f$ is convex. Therefore, $R^f(\cdot)$ is non-decreasing on $(1/2,1)$, and non-increasing on $(0,1/2)$. This means that it is $R^f(x)$ is minimized at $x = 1/2$. That $R^f(1/2) = f(1/2)$ immediately follows from definition of $R^f$.
\end{proof}

\begin{lemma} \label{lem:integral}
For any weakly proper scoring rule $f$:
\[\int_0^1 R(x) dx = f \parens{\half} + \int_\half^1 (1 - x) f'(x) dx = \half f \parens{\half} + \int_\half^1 f(x) dx.\]
\end{lemma}

\begin{proof}[Proof of Lemma~\ref{lem:integral}]
We have
\[\int_0^1 R(x) dx = \int_0^1 (xf(x) + (1 - x)f(1 - x)) dx = \int_0^1 2x f(x) dx\]
where the last step follows by separating $(1 - x)f(1 - x)$ into its own integral and substituting $u = 1 - x$. Now we integrate by parts, letting $u = f(x)$ and $dv = 2x dx$, so that $du = f'(x)$ and $v = x^2$, to get
\begin{align*}
    \int_0^1 R(x) dx &= x^2 f(x) \mid_0^1 - \int_0^1 x^2 f'(x) dx = f(1) - \int_0^1 x^2 f'(x) dx\\
    &= f(1) - \int_0^\half x^2 f'(x) dx - \int_\half^1 x^2 f'(x) dx\\
    &= f(1) - \int_0^\half x(1 - x)f'(1 - x) dx - \int_\half^1 x^2 f'(x) dx\\
    &= f(1) - \int_\half^1 (x(1 - x) + x^2)f'(x) dx = f(1) - \int_\half^1 xf'(x) dx.
\end{align*}
Expressing $f(1) - f \parens{\half}$ as $\int_\half^1 f'(x) dx$, we obtain the first equality. To obtain the equality with the third expression, we use integration by parts again, letting $u = x$ and $dv = f'(x) dx$, so that $du = dx$ and $v = f(x)$, to get that $\int_0^1 R(x) dx$ is equal to
\[f(1) - xf(x) \mid_\half^1 + \int_\half^1 f(x) dx = f(1) - f(1) + \half f \parens{\half} + \int_\half^1 f(x) dx = \half f \parens{\half} + \int_\half^1 f(x) dx.\]
\end{proof}

\begin{corollary}\label{cor:normalized} A scoring rule $f(\cdot)$ is \emph{normalized} if $\int_\half^1 (1-x)f'(x)dx = 1$, and $f(1/2) = 0$. 
\end{corollary}

\begin{proof}[Proof of Corollary~\ref{cor:normalized}]
Simply combine Definition~\ref{def:normalized} and Lemma~\ref{lem:integral}.
\end{proof}

\section{Properties of Respectful Scoring Rules}\label{app:respect}
We state several sufficient conditions for a scoring rule to be respectful, confirm that typical scoring rules are respectful, and provide a brief discussion.

\begin{claim} \label{claim:easy}
If Conditions 1 and 2 of Definition~\ref{def:respectful} hold and $\abs{R'''(x)}$ is bounded on $(0, 1)$ then $f$ is respectful.
\end{claim}

This should be clear: take $c$ to be small enough such that $c^{-.16}$ times the lower bound on $R''$ exceeds the upper bound on $\abs{R'''}$. Scoring rules such as the quadratic scoring rule and the spherical scoring rule satisfy the hypotheses of Claim~\ref{claim:easy}. One well-known generalization of the quadratic scoring rule is the Tsallis rule~\cite{Tsallis}. This rule, parametrized by $\gamma > 1$, is defined to be the scoring rule $f$ for which $R^f(x) = x^\gamma + (1 - x)^\gamma$. When $\gamma = 2$, this yields the quadratic scoring rule.

For $\gamma \ge 3$, it is evident that the Tsallis rule satisfies the hypotheses of Claim~\ref{claim:easy}. However, this is not so for $\gamma < 3$ (except for $\gamma = 2$). Perhaps more importantly, the logarithmic scoring rule does not satisfy Claim~\ref{claim:easy} either. This motivates the following result (the proof appears at the end of this section).

\begin{claim} \label{claim:interesting}
Suppose that Conditions 1 and 2 of Definition~\ref{def:respectful} are satisfied. Suppose further that $R'''$ is bounded on any closed sub-interval of $(0, 1)$, and that there exist constants $k \neq 0$ and $r$ such that $\lim_{x \to 0} x^r R'''(x) = k$. Then $f$ is respectful.
\end{claim}

The logarithmic scoring rule satisfies the hypotheses of Claim~\ref{claim:interesting} ($r = 2$ and $k = -1$). The Tsallis rule with $\gamma < 3$ (and $\gamma \neq 2$) also satisfies these hypotheses ($r = 3 - \gamma$ and $k = \gamma(\gamma - 1)(\gamma - 2)$). The proof of Claim~\ref{claim:interesting} is more involved, and deferred to Appendix~\ref{app:index}.


We briefly discuss ways in which proper scoring rules can fail to be respectful. One way a scoring rule can be disrespectful is if $R''$ grows extremely quickly near zero (e.g. $R''(x) = e^{1/x}$). Such functions, however, are outside the scope of this entire exercise because they are not normalizable. That is, such $R$ have $\int_0^1 R(x) dx = \infty$, and provide infinite expected payment to the expert. So this ``limitation'' of respectfulness is more of a restatement of normalizability.

Another way a scoring rule could be disrespectful is if $R''$ is not bounded away from zero. For example: $R''(x) = \parens{x - \frac{1}{2}}^2$ or $R''(x) = x(1 - x)$. If $R''$ remains ``very flat'' near $0$ for a ``large interval'', then $\ind^\ell(f)$ is infinite anyway. This implies that we should expect the error to be a \emph{larger} order of magnitude than $c^{-\ell/4}$, and for such functions to not incentivize precision well at all (although we do not explicitly prove this). This makes sense: if $R''(x) \approx 0$, then the expert gains $\approx 0$ by flipping the coin to refine their current belief (Lemma~\ref{lem:increward}). It is also possible that $R''$ is not bounded away from zero, but also not ``very flat''. We conjecture that Theorem~\ref{thm:global} does hold for such functions, but that our approach does not establish this. While it is possible to come up with such functions (e.g., the two above) which elude Theorem~\ref{thm:global}'s precise statement, this does not affect commonly-studied scoring rules, nor the scoring rules designed in this paper (sometimes leaning on Theorem~\ref{thm:approx}).

Finally, as with any exercise in analysis, there are continuous functions that behave erratically near zero, such as $R''(x) = \sin \parens{\frac{1}{x}} + \sin \parens{\frac{1}{1 - x}} + 3$. While it may or may not be the case that Theorem~\ref{thm:global} extends to such functions, this does not seem particularly relevant.

\begin{proof}[Proof of Claim~\ref{claim:interesting}]
Let $r, k$ be as in Claim~\ref{claim:interesting}. If $r \le 0$ then the claim is uninteresting: $R'''$ is bounded on $(0, 1)$ and so the statement is subsumed by Claim~\ref{claim:easy}. The interesting case is when $r > 0$.

We first consider the case when $r > 1$. Note that $\lim_{x \to 0} R''(x) = \infty$. To see this, suppose for contradiction that this limit is finite. We may write
\[\lim_{x \to 0} R''(x) = \lim_{x \to 0} \frac{xR''(x)}{x} = \lim_{x \to 0} xR'''(x) + R''(x)\]
by L'H\^{o}pital's rule, so $\lim_{x \to 0} xR'''(x) = 0$, contradicting that $\lim_{x \to 0} x^r R'''(x) \neq 0$.

Now, the fact that $\lim_{x \to 0} R''(x) = \infty$ lets us apply L'H\^{o}pital's rule:
\[\lim_{x \to 0} x^{r - 1} R''(x) = \lim_{x \to 0} \frac{R''(x)}{x^{1 - r}} = \lim_{x \to 0} \frac{R'''(x)}{(1 - r)x^{-r}} = \frac{1}{1 - r} \lim_{x \to 0} x^r R'''(x) = \frac{k}{1 - r}.\]
This means that
\[\lim_{x \to 0} x \frac{R'''(x)}{R''(x)} = 1 - r\]
so in particular, there exists $\delta > 0$ such that for all $x \le \delta$ we have $x \frac{R'''(x)}{R''(x)} \in [-r, 2 - r]$ and so $\frac{\abs{R'''(x)}}{R''(x)} \le \frac{r}{x}$. On the other hand, $R'''$ is bounded on $[\delta, 1 - \delta]$ by assumption.

To finish, let $t$ as in Definition~\ref{def:respectful} equal $0.3$. Assume $c$ is small enough that the following conditions are satisfied:
\begin{itemize}
    \item $\frac{\abs{R'''(x)}}{R''(x)} \le c^{-.16}$ on $[\delta, 1 - \delta]$.
    \item $c^{-.01} \ge r$.
\end{itemize}
Then the last condition of Definition~\ref{def:respectful} will be satisfied on $[\delta, 1 - \delta]$; it will also be satisfied on $[c^{.3}, \delta]$ for any $c$ satisfying the second condition above, because on that interval we have

\[\frac{\abs{R'''(x)}}{R''(x)} \le \frac{r}{x} \le \frac{1}{c^{.16} \sqrt{x(1 - x)}} \cdot \frac{c^{.16}r}{\sqrt{x}} \le \frac{1}{c^{.16} \sqrt{x(1 - x)}} \cdot \frac{c^{.16}r}{c^{.15}} \le \frac{1}{c^{.16} \sqrt{x(1 - x)}}.\]

By symmetry of $R''$ about $\frac{1}{2}$ (and antisymmetry of $R'''$) we have that the condition also holds on $[1 - \delta, 1 - c^t]$, as desired.

Now we consider the case that $r = 1$. As above, we have $\lim_{x \to 0} R''(x) = \infty$. Proceeding similarly, we have
\[\lim_{x \to 0} \frac{R''(x)}{\ln x} = \lim_{x \to 0} \frac{R'''(x)}{\frac{1}{x}} = \lim_{x \to 0} xR'''(x) = k.\]
This means that there exists $\delta > 0$ such that for all $x \le \delta$ we have $x \ln x \frac{R'''(x)}{R''(x)} \in [0, 2]$ and so $\frac{\abs{R'''(x)}}{R''(x)} \le \frac{-2}{x \ln x}$. We finish as before.

Finally, consider the case that $0 < r < 1$. Let $a$ be a lower bound on $R''$, as in the statement of Claim~\ref{claim:interesting}. It suffices to show that for for $c$ small enough, we have $\abs{R'''(x)} \le \frac{a}{c^{.16}\sqrt{x}}$ on $[c^{.3}, 1 - c^{.3}]$. Let $\delta$ be such that $x^r R'''(x) \in [k - 1, k + 1]$ for all $x \le \delta$. On $[c^{.3}, \delta]$ we have
\[\abs{R'''(x)} \le \frac{\abs{k} + 1}{x^r} \le \frac{\abs{k} + 1}{x} = \frac{a}{c^{.16} \sqrt{x}} \cdot \frac{\abs{k} + 1}{ac^{-.16} \sqrt{x}} \le \frac{a}{c^{.16} \sqrt{x}} \cdot \frac{\abs{k} + 1}{ac^{-.16} \cdot c^{.15}}\]
if $c$ is small enough that $c^{-.01} \ge \frac{\abs{k} + 1}{a}$. (As before, we also need to make sure that $c$ is small enough that the condition is satisfied on $[\delta, 1 - \delta]$.) This concludes the proof.
\end{proof}

\section{Omitted Proofs from Section~\ref{sec:index}}\label{app:index}

\begin{proof}[Proof of Claim~\ref{claim:rdoubleprime}]
Say $h$ of the first $n$ flips were heads, so $q = \frac{h + 1}{n + 2}$. The expert's expected reward is $R \parens{\frac{h + 1}{n + 2}}$. The expert reasons: with probability $q$, the next coin will come up heads and my new estimate will be $\frac{h + 2}{n + 3}$; with probability $1 - q$ it will be come up tails and my new estimate will be $\frac{h + 1}{n + 3}$. Therefore, the expert's expected increase in reward from flipping the $n + 1$-th coin is
\[\Delta_{n + 1} = \frac{h + 1}{n + 2} R \parens{\frac{h + 2}{n + 3}} + \frac{n - h + 1}{n + 2} R \parens{\frac{h + 1}{n + 3}} - R \parens{\frac{h + 1}{n + 2}}.\]
Since $R$ is twice differentiable, we may use Taylor's approximation theorem to write
\[R \parens{\frac{h + 1}{n + 3}} = R \parens{\frac{h + 1}{n + 2}} + \parens{\frac{h + 1}{n + 3} - \frac{h + 1}{n + 2}} R' \parens{\frac{h + 1}{n + 2}} + \frac{1}{2} \parens{\frac{h + 1}{n + 3} - \frac{h + 1}{n + 2}}^2 R''(c_1)\]
for some $c_1 \in \brackets{\frac{h + 1}{n + 3}, \frac{h + 1}{n + 2}}$. Similarly we have
\[R \parens{\frac{h + 2}{n + 3}} = R \parens{\frac{h + 1}{n + 2}} + \parens{\frac{h + 2}{n + 3} - \frac{h + 1}{n + 2}} R' \parens{\frac{h + 1}{n + 2}} + \frac{1}{2} \parens{\frac{h + 2}{n + 3} - \frac{h + 1}{n + 2}}^2 R''(c_1)\]
for some $c_2 \in \brackets{\frac{h + 1}{n + 2}, \frac{h + 2}{n + 3}}$. When we plug these expressions into the formula for $\Delta_{n + 1}$ above, the zeroth- and first-order terms cancel. We are left with
\begin{align*}
\Delta_{n + 1} &= \frac{n - h + 1}{n + 2} \cdot \frac{1}{2} \parens{\frac{h + 1}{n + 3} - \frac{h + 1}{n + 2}}^2 R''(c_1) + \frac{h + 1}{n + 2} \cdot \frac{1}{2} \parens{\frac{h + 2}{n + 3} - \frac{h + 1}{n + 2}}^2 R''(c_2)\\
&= \frac{(h + 1)^2(n - h + 1)}{2(n + 2)^3(n + 3)^2} R''(c_1) + \frac{(h + 1)(n - h + 1)^2}{2(n + 2)^3(n + 3)^2} R''(c_2)\\
&= \frac{q(1 - q)}{2(n + 3)^2}(qR''(c_1) + (1 - q)R''(c_2)).
\end{align*}
Note that $\abs{c_1 - q} \le \frac{h + 1}{n + 2} - \frac{h + 1}{n + 3} \le \frac{1}{n}$, so $c_1 \in [q - \frac{1}{n}, q + \frac{1}{n}]$, and similarly for $c_2$. This completes the proof.
\end{proof}

\begin{proof}[Proof of Claim~\ref{claim:one_third_bound}]
Suppose that $R''(x) \ge a$ for all $x \in (0, 1)$. By Claim~\ref{claim:rdoubleprime} we have
\[\Delta_{n + 1} \ge \frac{q(1 - q)a}{2(n + 3)^2}.\]
Now, we have that $\frac{1}{n + 2} \le q \le \frac{n - 1}{n + 2}$, and $q(1 - q)$ decreases as $q$ gets farther from $\frac{1}{2}$. This means that
\[\Delta_{n + 1} \ge \frac{\frac{n + 1}{(n + 2)^2} a}{2(n + 3)^2} \ge \frac{a}{72(n + 1)^3}.\]
Therefore, if $\Delta_{n + 1} < c$ then $n + 1 > \frac{1}{\parens{\frac{72}{a}}^{1/3}c^{1/3}}$, so $n > \frac{1}{\alpha c^{1/3}}$ for some $\alpha$ (not to be confused with $a$), if $c$ is small enough.
\end{proof}

\begin{proof}[Proof of Claim~\ref{claim:omega_p}]
Let $1 \le j_p \le n - 2$ be such that $\frac{j_p}{n} \le p \le \frac{j_p + 1}{n}$. We have
\[\frac{j_p}{n} - \frac{\sqrt{j_p(n - j_p)}}{2n^{1.49}} \le Q_{j_p/n}(n) \le Q_p(n) \le Q_{(j_p + 1)/n}(n) \le \frac{j_p + 1}{n} + \frac{\sqrt{(j_p + 1)(n - 1 - j_p)}}{2n^{1.49}}\]
so
\begin{align*}
\abs{Q_p(n) - p} &\le \max \parens{p - \frac{j_p}{n} + \frac{\sqrt{j_p(n - j_p)}}{2n^{1.49}}, \frac{j_p + 1}{n} - p + \frac{\sqrt{(j_p + 1)(n - 1 - j_p)}}{2n^{1.49}}}\\
&\le \frac{1}{n} + \frac{1}{2n^{1.49}} \max \parens{\sqrt{j_p(n - j_p)}, \sqrt{(j_p + 1)(n - 1 - j_p)}}.
\end{align*}
For fixed $n$ and for $\frac{1}{n} \le p \le 1 - \frac{1}{n}$, this maximum divided by $n\sqrt{p(1 - p)}$ is maximized when $p = \frac{1}{n}$ (or $p = 1 - \frac{1}{n}$), in which case the ratio is $\sqrt{\frac{2(n - 2)}{n - 1}} \le \sqrt{2}$. Therefore we have
\[\frac{\abs{Q_p(n) - p}}{\sqrt{p(1 - p)}} \le \frac{1}{n\sqrt{p(1 - p)}} + \frac{n\sqrt{2}}{2n^{1.49}} \le \frac{1}{n^{.49}}\]
for $n$ large enough. (Here we again use that $p \ge \frac{1}{n}$, so $\sqrt{p(1 - p)}$ is minimized at $p = \frac{1}{n}$.)
\end{proof}

\begin{proof}[Proof of Claim~\ref{claim:omega_unlikely}]
We have
\[\pr{\overline{\Omega_N}} \le \sum_{n = N}^\infty \sum_{j = 1}^{n - 1} \pr{\abs{Q_{j/n}(n) - \frac{j}{n}} > \frac{\sqrt{j(n - j)}}{2n^{1.49}}}.\]
Now, let $G_{j/n}(n)$ be the fraction of the first $n$ coin flips that were heads (so $G_{j/n}(n)$ is an average of $n$ i.i.d. Bernoulli random variables that are $1$ with probability $\frac{j}{n}$). Note that $Q_{j/n}(n)$ is within $\frac{1}{n}$ of $G_{j/n}(n)$, and for large $n$ we have $\frac{1}{n} \le \frac{\sqrt{j(n - j)}}{2n^{1.49}}$ for all $j$. This means that for large $n$, by the triangle inequality we have that if $\abs{G_{j/n}(n) - \frac{j}{n}} > \frac{\sqrt{j(n - j)}}{n^{1.49}}$ then $\abs{Q_{j/n}(n) - \frac{j}{n}} > \frac{\sqrt{j(n - j)}}{2n^{1.49}}$. Therefore, for large $N$ we have
\[\pr{\overline{\Omega_N}} \le \sum_{n = N}^\infty \sum_{j = 1}^{n - 1} \pr{\abs{G_{j/n}(n) - \frac{j}{n}} > \frac{\sqrt{j(n - j)}}{n^{1.49}}}.\]
We bound each of these probabilities. Recall the following version of the Chernoff bound: for $0 \le \delta \le 1$, if $X = \sum_{i = 1}^n X_i$ is a sum of i.i.d. Bernoulli random variables with $\EE{X} = \mu$, then
\[\pr{\abs{X - \mu} \ge \delta \mu} \le 2e^{-\mu \delta^2/3}.\]
We apply this to our random variables (so $X = nG_{j/n}(n)$ and $\mu = j$). Assume $j \le \frac{n}{2}$. Let $\delta = n^{-.49} \sqrt{\frac{n - j}{j}}$. Then
\[\pr{\abs{\frac{X}{n} - \frac{j}{n}} \ge \frac{\sqrt{j(n - j)}}{n^{-1.49}}} \le 2e^{-n^{.02}/6}.\]
If $j \ge \frac{n}{2}$, a symmetry argument yields the same result. Therefore, for sufficiently large $N$ we have
\begin{align*}
\pr{\overline{\Omega_N}} &\le \sum_{n = N}^\infty \sum_{j = 1}^{n - 1} 2e^{-n^{.02}/6} \le \sum_{n = N}^\infty 2ne^{-n^{.02}/6} \le \sum_{n = N}^\infty e^{-n^{.02}/7} = e^{-N^{.02}/7} \sum_{n = N}^\infty e^{-(n^{.02} - N^{.02})/7}\\
&\le e^{-N^{.02}/7} \sum_{n = N}^\infty e^{-n^{.02}/14} \le e^{-N^{.02}/7} \sum_{n = N}^\infty \frac{14^{100} \cdot 100!}{n^2}.
\end{align*}
The last step comes from observing that $e^x \ge \frac{x^{100}}{100!}$ for positive $x$ and plugging in $x = \frac{n^{.02}}{14}$. Now, this summand is bounded by a constant, since $\sum_{n = 1}^\infty \frac{1}{n^2}$ converges, and so we have
\[\pr{\overline{\Omega_N}} \le O \parens{e^{-N^{.02}/7}} \le O \parens{e^{-N^{.01}}},\]
as desired.
\end{proof}

\begin{proof}[Proof of Proposition~\ref{prop:n_stop_bound}]
We assume for convenience that $t < 0.3$ (which is safe, as Definition~\ref{def:respectful} holds for all $t \leq t'$ whenever it holds for $t'$).

Fix $p \in [2c^t, 1 - 2c^t]$. Let $\alpha$ be as in Claim~\ref{claim:one_third_bound}. As before, let $Q(n)$ be the predictor's estimate for the bias of the coin after $n$ flips. Then for $c$ small enough that $\frac{1}{\alpha c^{1/3}} \ge N$ and $c^{1/30} \le \frac{2}{\alpha}$, for $n \ge \frac{1}{\alpha c^{1/3}}$, we have
\[p - \sqrt{p(1 - p)}(\alpha c^{1/3})^{.49} \le Q(n) \le p + \sqrt{p(1 - p)}(\alpha c^{1/3})^{.49}.\]
This follows from Claim~\ref{claim:omega_p}, noting that if $c^{1/30} \le \frac{2}{\alpha}$ and $n \ge \frac{1}{\alpha c^{1/3}}$ then $\frac{1}{n} \le p \le 1 - \frac{1}{n}$.

Now, recall Claim~\ref{claim:rdoubleprime}:
\[\Delta_{n + 1} = \frac{Q(n)(1 - Q(n))}{2(n + 3)^2}(Q(n)R''(c_1) + (1 - Q(n))R''(c_2))\]
for some $c_1, c_2 \in [Q(n) - \frac{1}{n}, Q(n) + \frac{1}{n}]$. In the remainder of this proof, what we essentially argue is that $R''$ on this interval is not too far from $R''(p)$, because of our bound on $Q(n)$ as $p$ plus or minus a small quantity.

We ask: for a given (possibly negative) $\epsilon$, how far from $R''(p)$ can $R''(p + \epsilon)$ be? Well, since $R'''$ is integrable, we have
\[\abs{R''(p + \epsilon) - R''(p)} = \abs{\int_p^{p + \epsilon} R'''(x) dx} \le \abs{\int_p^{p + \epsilon} \abs{R'''(x)} dx}.\]
Now, since $f$ is respectful we have that for $c$ small enough, if $p, p + \epsilon \in [c^t, 1 - c^t]$ then
\[\abs{R'''(x)} \le \frac{1}{c^{.16}\sqrt{x(1 - x)}} R''(x) \le \frac{1}{c^{.16}\sqrt{\hat{p}(1 - \hat{p})}} R''(x),\]
where $\hat{p}$ is defined to be the number on the interval between $p$ and $p + \epsilon$ minimizing $\sqrt{x(1 - x)}$ (i.e. farthest from $\frac{1}{2}$).

Define $r := \frac{1}{c^{.16}\sqrt{\hat{p}(1 - \hat{p})}}$. Then $\abs{R'''(x)} \le rR''(x)$. We use this fact to prove the following.

\begin{claim} \label{claim:diff_ineq}
\[\abs{R''(p + \epsilon) - R''(p)} \le R''(p)(e^{r\abs{\epsilon}} - 1).\]
\end{claim}

\begin{proof}
We prove this for positive $\epsilon$. The result follows for negative $\epsilon$ because if some $R = R_1$ is a counterexample for some negative $\epsilon_1$, then $R = R_2$ where $R_2'''(p + x) := -R_1'''(p - x)$ for $x \in [0, \epsilon_2]$ serves as a counterexample for $\epsilon_2 := -\epsilon_1$. (This is because $R_2''(p + x) = R_1''(p - x)$ for any $x \in [0, \epsilon_2]$, by the fundamental theorem of calculus.) Additionally, we may assume that $R''(p) = 1$, because if there is a counterexample function $R = R_1$ to the claim then $R = R_2$ where $R_2(x) = \frac{R_1(x)}{R''(p)}$ also serves as a counterexample.

We prove that
\[e^{-r\epsilon} - 1 \le R''(p + \epsilon) - 1 \le e^{r\epsilon} - 1.\]
The left inequality suffices because $1 - e^{-x} \le e^x - 1$ for all $x$, so in particular $1 - e^{r\epsilon} \le e^{-r\epsilon} - 1$.

We begin with the right inequality. Suppose for contradiction that $R''(p + \epsilon) > e^{r\epsilon}$. Let $S$ be the set of points in $[p, p + \epsilon]$ where $R''(x) > e^{r(x - p)}$. Since $S$ contains $p + \epsilon$, it is nonempty; let $p_1 = \inf_S$. Since $R''$ is continuous, we have $R''(p_1) - e^{r(p_1 - p)} = 0$. Pick $\delta > 0$ small enough that the set $T$ of points $x \in \brackets{p_1, p_1 + \min(\epsilon, \frac{1}{3r})}$ with $R''(x) - e^{r(x - p)} > \delta$ is nonempty. Let $p_2 = \inf_T$, so $R''(p_2) - e^{r(p_2 - p)} = \delta$. Note that
\[\delta = R''(p_2) - e^{r(p_2 - p)} = R''(p_1) - e^{r(p_1 - p)} + \int_{p_1}^{p_2} \frac{d}{dx} (R''(x) - e^{r(x - p)}) dx = \int_{p_1}^{p_2} (R'''(x) - re^{r(x - p)}) dx.\]
It follows that $R'''(p_3) - re^{r(p_3 - p)} \ge \frac{\delta}{p_2 - p_1} \ge 2r\delta$ for some $p_3 \in [p_1, p_2]$. (Otherwise the value of the integral would be at most the integral of $2r\delta$ from $p_1$ to $p_2$, which is at most $\frac{2}{3} \delta$, since $p_2 - p_1 \le \frac{1}{3r}$.) Therefore, because $\abs{R'''(x)} \le rR''(x)$ for all $x \in [p, p + \epsilon]$, we have that
\[rR''(p_3) - re^{r(p_3 - p)} \ge 2r\delta\]
so $R''(p_3) - e^{r(p_3 - p)} \ge 2\delta$. But then we have that $p_2 < p_3$ and $p_3 \in T$, contradicting the definition of $p_2$ as the infimum of $T$.

The proof of the left inequality above proceeds similarly, but is not exactly analogous. Suppose for contradiction that $t := R''(p + \epsilon) - e^{-r\epsilon} < 0$. Define $p_1$ to be the supremum of points in $[p, p + \epsilon]$ where $R''(x) \ge e^{-r(x - p)}$ (so $R''(p_1) = e^{-r(p_1 - p)}$). Then $R''(x) - e^{-r(x - p)}$ is zero at $x = p_1$ and $t$ at $x = p + \epsilon$, so
\[\int_{p_1}^{p + \epsilon} (R'''(x) + re^{-r(x - p)}) dx = t < 0,\]
which means that for some $p_2 \in [p_1, p + \epsilon]$ we have that $R'''(p_2) + re^{-r(p_2 - p)} \le \frac{t}{2(p + \epsilon - p_1)}$ (otherwise the value of the integral would be at least $\frac{t}{2}$). Since $\abs{R'''(x)} \le rR''(x)$ for all $x \in [p, p + \epsilon]$, we have that
\[-rR''(p_2) + re^{-r(p_2 - p)} \le \frac{t}{2(p + \epsilon - p_1)} < 0,\]
so $R''(p_2) > e^{-r(p_2 - p)}$. This is a contradiction, since on the one hand we have $p_2 > p_1$, but on the other hand $p_1$ was defined as the supremum of points where $R''(x) \ge e^{-r(x - p)}$. This completes the proof.
\end{proof}

How large of an $\epsilon$ do we care about? The farthest that $c_1$ and $c_2$ can be from $p$ is $\sqrt{p(1 - p)}(\alpha c^{1/3})^{.49} + \alpha c^{1/3} \le 2\sqrt{p(1 - p)}(\alpha c^{1/3})^{.49}$, for small $c$. (This is because we assumed for convenience that $t < 0.3$, which means that $p(1 - p) \ge c^{.3}$, so $\alpha c^{1/3} \le \sqrt{p(1 - p)}(\alpha c^{1/3})^{.49}$.) Therefore, by Claim~\ref{claim:diff_ineq} we have
\begin{align*}
\abs{R''(p) - R''(c_1)}, \abs{R''(p) - R''(c_2)} &\le R''(p)(e^{2r\sqrt{p(1 - p)}(\alpha c^{1/3})^{.49}} - 1)\\
&= R''(p) \exp \parens{\frac{2\sqrt{p(1 - p)}(\alpha c^{1/3})^{.49}}{c^{.16}\sqrt{\hat{p}(1 - \hat{p})}}} - R''(p)
\end{align*}
where $\hat{p}$ is either $p \pm 2\sqrt{p(1 - p)}(\alpha c^{1/3})^{.49}$, whichever is farther from $\frac{1}{2}$. It is easy to check\footnote{Without loss of generality assume $p \le \frac{1}{2}$, so $\hat{p} = p - 2\sqrt{p(1 - p)}(\alpha c^{1/3})^{.49}$. Then $\frac{\sqrt{p(1 - p)}}{\sqrt{\hat{p}(1 - \hat{p})}} \le \frac{p(1 - p)}{\hat{p}(1 - \hat{p})} \le \frac{p}{\hat{p}}$, so it suffices to show that $2\sqrt{p(1 - p)}(\alpha c^{1/3})^{.49} \le p$. This is indeed the case, as $p \ge c^{.3} \ge 4(\alpha c^{1/3})^{.98}$ for small $c$, so $2\sqrt{p}(\alpha c^{1/3})^{.49} \le p$.} that for small enough $c$ we have that $\frac{\sqrt{p(1 - p)}}{\sqrt{\hat{p}(1 - \hat{p})}} \le 2$, and so we have
\[\abs{R''(p) - R''(c_1)}, \abs{R''(p) - R''(c_2)} \le R''(p) \exp \parens{\frac{1}{2} \beta c^{1/300}} - R''(p) \le \beta c^{1/300} R''(p)\]
for small enough $c$, where $\beta = 8\alpha^{.49}$. (Here we use that $e^x \le 1 + 2x$ for small positive $x$.) It follows, then, by Claim~\ref{claim:rdoubleprime}, that
\[\frac{Q(n)(1 - Q(n))}{2(n + 3)^2} R''(p)(1 - \beta c^{1/300}) \le \Delta_{n + 1} \le \frac{Q(n)(1 - Q(n))}{2(n + 3)^2} R''(p)(1 + \beta c^{1/300}).\]
Note that since $Q(n) \ge p - \sqrt{p(1 - p)}(\alpha c^{1/3})^{.49}$ and $1 - Q(n) \ge 1 - p - \sqrt{p(1 - p)}(\alpha c^{1/3})^{.49}$, we may write
\begin{align*}
Q(n)(1 - Q(n)) &\ge p(1 - p)(1 + (\alpha c^{1/3})^{.98}) - \sqrt{p(1 - p)}(\alpha c^{1/3})^{.49} = p(1 - p) \parens{1 - \frac{(\alpha c^{1/3})^{.49}}{\sqrt{p(1 - p)}}}\\
&\ge p(1 - p) \parens{1 - \frac{2(\alpha c^{1/3})^{.49}}{c^{t/2}}} \ge p(1 - p)(1 - \alpha^{.49} c^{.01})
\end{align*}
for $c$ small enough that the second-to-last step holds. (In the last step we use that $t < 0.3$.) A similar calculation shows that $Q(n)(1 - Q(n)) \le p(1 - p)(1 + \alpha^{.49} c^{.01})$ for $c$ small enough.\footnote{An extra $\alpha^{.49} c^{.49/3}$ appears, but this term is dominated by $\alpha^{.49} c^{.01}$ for small $c$.} Also note that $n^2 \le (n + 3)^2 \le n^2(1 + 4\alpha c^{1/3})^2$. Putting these approximations all together, we note that the $c^{1/300}$ approximation is the dominant one, which means that there is a constant $\gamma$ such that
\begin{equation} \label{eq:delta_bound}
\frac{p(1 - p)}{2n^2}R''(p)(1 - \gamma c^{1/300}) \le \Delta_{n + 1} \le \frac{p(1 - p)}{2n^2}R''(p)(1 + \gamma c^{1/300}).
\end{equation}
Therefore, since the expert stops flipping when $\Delta_{n + 1} < c$, we have
\[\sqrt{\frac{p(1 - p)R''(p)}{2c}(1 - \gamma c^{1/300})} \le \nstop \le \sqrt{\frac{p(1 - p)R''(p)}{2c}(1 + \gamma c^{1/300})}.\]
This holds for any $p$ such that $p \pm 2\sqrt{p(1 - p)}(\alpha c^{1/3})^{.49} \in [c^t, 1 - c^t]$; a sufficient condition is $p \in [2c^t, 1 - 2c^t]$.
\end{proof}

\emph{A note on terminology:} We will sometimes say that a function $g(c)$ is $o(h(c))$ \emph{uniformly} over $p$. This means that $g$ and $h$ are implicitly functions of $p$ as well, but that $\frac{h(x)}{g(x)}$ approaches zero uniformly in $p$ (i.e. for all $\epsilon$ there exists $c_\epsilon$ such that for all $c \le c_\epsilon$, we have $\frac{h(x)}{g(x)} < \epsilon$ for all (relevant) values of $p$). So for instance, the $o(1)$ in the statement of Lemma~\ref{lem:close} is uniform in $p$.

To prove Lemma~\ref{lem:close} we first prove the following general proposition.

\begin{prop} \label{prop:exp_l}
Let $X_{c,p}$ and $Y_{c,p}$ be random variables taking values in $[0, 1]$ for each real number $c > 0$ and $p \in \mathcal{P}_c$ (some arbitrary set that depends on $c$). Let $\ell > 0$. If $\lim_{c \to 0} \frac{\EE{Y_{c,p}^\ell}}{\EE{X_{c,p}^\ell}} = 0$ uniformly over $p \in \mathcal{P}_c$, then $\lim_{c \to 0} \frac{\EE{(X_{c,p} + Y_{c,p})^\ell}}{\EE{X_{c,p}^\ell}} = 1$ uniformly over $p \in \mathcal{P}_c$. Separately, if $\lim_{c \to 0} \frac{\EE{Y_{c,p}^\ell}}{\EE{(X_{c,p} + Y_{c,p})^\ell}} = 0$ uniformly over $p \in \mathcal{P}_c$, then $\lim_{c \to 0} \frac{\EE{X_{c,p}^\ell}}{\EE{(X_{c,p} + Y_{c,p})^\ell}} = 1$ uniformly over $p \in \mathcal{P}_c$.
\end{prop}

\begin{proof}
The condition that $X_{c,p}, Y_{c,p} \in [0, 1]$ is simply a convenient one to guarantee that all relevant expectations are finite. Now, for any $a \in (0, 1)$, we have
\begin{align} \label{eq:exp_l}
\EE{(X_{c,p}+Y_{c,p})^\ell} &= \int_0^\infty \pr{(X_{c,p}+Y_{c,p})^\ell > z}dz = \int_0^\infty \pr{X_{c,p}+Y_{c,p} > z^{1/\ell}}dz \nonumber\\
&\leq \int_0^\infty \parens{\pr{X_{c,p} > (1-a)z^{1/\ell}} + \pr{Y_{c,p} > az^{1/\ell}}}dz \nonumber\\
&= \int_0^\infty \parens{\pr{\frac{X_{c,p}^\ell}{(1-a)^\ell} > z} + \pr{\frac{Y_{c,p}^\ell}{a^\ell} > z}} dz = \frac{\EE{X_{c,p}^\ell}}{(1-a)^\ell} + \frac{\EE{Y_{c,p}^\ell}}{a^\ell},
\end{align}
where the inequality follows by a union bound.

We start with the first statement. Dividing Equation~\ref{eq:exp_l} by $\EE{X_{c,p}^\ell}$, we have
\[1 \le \frac{\EE{(X_{c,p} + Y_{c,p})^\ell}}{\EE{X_{c,p}^\ell}} \le \frac{1}{(1 - a)^\ell} + \frac{\EE{Y_{c,p}^\ell}}{\EE{X_{c,p}^\ell}} \cdot \frac{1}{a^\ell}.\]
The limit of $\frac{\EE{Y_{c,p}^\ell}}{\EE{X_{c,p}^\ell}}$ as $c$ approaches zero is $0$ by assumption, so for $c$ small enough we have that $\frac{\EE{Y_{c,p}^\ell}}{\EE{X_{c,p}^\ell}} \cdot \frac{1}{a^\ell} \le a$ for all $p \in \mathcal{P}_c$. In other words, for every $a$ there exists $c_a$ such that for all $c \le c_a$ we have that $1 \le \frac{\EE{(X_{c,p} + Y_{c,p})^\ell}}{\EE{X_{c,p}^\ell}} \le \frac{1}{(1 - a)^\ell} + a$. Since $\lim_{a \to 0} \frac{1}{(1 - a)^\ell} + a = 1$, we have that for all $\epsilon > 0$ there exists $c_\epsilon$ such that for all $c \le c_\epsilon$ and $p \in \mathcal{P}_c$ we have that $1 \le \frac{\EE{(X_{c,p} + Y_{c,p})^\ell}}{\EE{X_{c,p}^\ell}} \le 1 + \epsilon$. This proves the first statement.

As for the second statement, we divide Equation~\ref{eq:exp_l} by $\EE{(X_{c,p} + Y_{c,p})^\ell}$ to obtain
\[1 \le \frac{1}{(1 - a)^\ell} \cdot \frac{\EE{X_{c,p}^\ell}}{\EE{(X_{c,p} + Y_{c,p})^\ell}} + \frac{1}{a^\ell} \cdot \frac{\EE{Y_{c,p}^\ell}}{\EE{(X_{c,p} + Y_{c,p})^\ell}}\]
so
\[(1 - a)^\ell\parens{1 - \frac{1}{a^\ell} \cdot \frac{\EE{Y_{c,p}^\ell}}{\EE{(X_{c,p} + Y_{c,p})^\ell}}} \le \frac{\EE{X_{c,p}^\ell}}{\EE{(X_{c,p} + Y_{c,p})^\ell}} \le 1.\]
The limit of $\frac{\EE{Y_{c,p}^\ell}}{\EE{(X_{c,p} + Y_{c,p})^\ell}}$ as $c$ approaches zero is $0$ by assumption, so for $c$ small enough we have that $\frac{1}{a^\ell} \cdot \frac{\EE{Y_{c,p}^\ell}}{\EE{(X_{c,p} + Y_{c,p})^\ell}} \le a$ for all $p \in \mathcal{P}_c$. In other words, for every $a$ there exists $c_a$ such that for all $c \le c_a$ we have that $(1 - a)^{\ell + 1} \le \frac{\EE{X_{c,p}^\ell}}{\EE{(X_{c,p} + Y_{c,p})^\ell}} \le 1$. Since $\lim_{a \to 0} (1 - a)^{\ell + 1} = 1$, we have that for all $\epsilon > 0$ there exists $c_\epsilon$ such that for all $c \le c_\epsilon$ and $p \in \mathcal{P}_c$ we have that $1 - \epsilon \le \frac{\EE{X_{c,p}^\ell}}{\EE{(X_{c,p} + Y_{c,p})^\ell}} \le 1$. This proves the second statement.
\end{proof}

We now prove Lemma~\ref{lem:close}.

\begin{proof}[Proof (of Lemma~\ref{lem:close})]
Fix any $c$ and $p \in [2c^t, 1 - 2c^t]$.
\[n_0 := \sqrt{\frac{p(1 - p)R''(p)}{2c}(1 - \gamma c^{1/300})}.\]
Let $d_1$ be the expert's error after $n_0$ flips, i.e. $\abs{Q(n_0) - p}$. Let $d_2$ be the distance from their guess after $n_0$ flips to their guess after $\nstop$ flips. Then the expert's error after $\nstop$ flips lies between $d_1 - d_2$ and $d_1 + d_2$ by the triangle inequality. That is, we have
\[\max(0, d_1 - d_2) \le \err_c(p) \le d_1 + d_2,\]
so
\[\EE{\max(0, d_1 - d_2)^\ell \mid \Omega_N} \le \EE{(\err_c(p))^\ell \mid \Omega_N} \le \EE{(d_1 + d_2)^\ell \mid \Omega_N}\]
which means that
\[\frac{\EE{\max(0, d_1 - d_2)^\ell \mid \Omega_N}}{\EE{d_1^\ell \mid \Omega_N}} \le \frac{\EE{(\err_c(p))^\ell \mid \Omega_N}}{\EE{d_1^\ell \mid \Omega_N}} \le \frac{\EE{(d_1 + d_2)^\ell \mid \Omega_N}}{\EE{d_1^\ell \mid \Omega_N}}.\]
We will later prove the following claim:
\begin{claim} \label{claim:d2_o_d1}
We have
\[\lim_{c \to 0} \frac{\EE{d_2^\ell \mid \Omega_N}}{\EE{d_1^\ell \mid \Omega_N}} = 0\]
uniformly over $p \in [2c^t, 1 - 2c^t]$. That is, for all $\epsilon$ there exists $c_\epsilon$ such that for all $c < c_\epsilon$, the fraction above is less than $\epsilon$ for all $p \in [2c^t, 1 - 2c^t]$.
\end{claim}
Now, given this claim, observe that by the first statement of Proposition~\ref{prop:exp_l} (with $X_{c,p} = d_1$ and $Y_{c,p} = d_2$) we have
\[\lim_{c \to 0} \frac{\EE{(d_1 + d_2)^\ell \mid \Omega_N}}{\EE{d_1^\ell \mid \Omega_N}} = 1.\]
By the second statement of the claim (with $X_{c,p} = \max(0, d_1 - d_2)$ and $Y_{c,p} = d_1 - \max(0, d_1 - d_2)$) we have
\[\lim_{c \to 0} \frac{\EE{\max(0, d_1 - d_2)^\ell \mid \Omega_N}}{\EE{d_1^\ell \mid \Omega_N}} = 1.\]
Note that the premise of the second claim holds for these $X_{c,p}$ and $Y_{c,p}$, because $Y_{c,p} = d_1 - \max(0, d_1 - d_2) \le d_2$ and so certainly if $\lim_{c \to 0} \frac{\EE{d_2^\ell \mid \Omega_N}}{\EE{d_1^\ell \mid \Omega_N}} = 0$ then $\lim_{c \to 0} \frac{\EE{(d_1 - \max(0, d_1 - d_2))^\ell \mid \Omega_N}}{\EE{d_1^\ell \mid \Omega_N}} = 0$.

By the squeeze theorem, it follows that
\[\lim_{c \to 0} \frac{\EE{(\err_c(p))^\ell \mid \Omega_N}}{\EE{d_1^\ell \mid \Omega_N}} = 1.\]

Note that this limit holds uniformly over $p \in [2c^t, 1 - 2c^t]$.

To complete the proof, we use the fact (proven in Claim~\ref{claim:be}) that
\[\EE{d_1^\ell \mid \Omega_N} = \mu_\ell \parens{\frac{p(1 - p)}{n_0}}^{\ell/2}(1 + o(1)).\]
We have
\[n_0 = \sqrt{\frac{p(1 - p)R''(p)}{2c}(1 - \gamma c^{1/300})} = \sqrt{\frac{p(1 - p)R''(p)}{2c}}(1 + o(1)).\]
Therefore we have
\[\EE{d_1^\ell \mid \Omega_N} = \mu_\ell \parens{\frac{2c}{p(1 - p)R''(p)}}^{\ell/4}(1 + o(1),\]
as desired.
\end{proof}

We now prove Claim~\ref{claim:d2_o_d1} by approximately computing $\EE{d_1^\ell \mid \Omega_N}$ and $\EE{d_2^\ell \mid \Omega_N}$. We begin with the former.

\begin{claim} \label{claim:be}
\[\EE{d_1^\ell \mid \Omega_N} = \mu_\ell \parens{\frac{p(1 - p)}{n_0}}^{\ell/2}(1 + o(1))\]
where the $o(1)$ term is a function of $c$ (but not $p$) that approaches zero as $c$ approaches zero.
\end{claim}

\begin{proof}
We use the Berry-Esseen theorem, a result about the speed of convergence of a sum of i.i.d. random variables to a normal distribution.

\begin{theorem}[Berry-Esseen theorem]
Let $X_1, \dots, X_n$ be i.i.d. random variables with $\EE{X_1} = 0$, $\EE{X_1^2} \equiv \sigma^2 > 0$, and $\EE{\abs{X_1}^3} = \rho < \infty$. Let $Y = \frac{1}{n} \sum_i X_i$ and let $F$ be the CDF of $\frac{Y\sqrt{n}}{\sigma}$. Let $\Phi(x)$ be the standard normal distribution. Then for all $x$ we have
\[\abs{F(x) - \Phi(x)} \le \frac{C \rho}{\sigma^3 \sqrt{n}},\]
for some universal constant $C$ independent of $n$ and the distribution of the $X_i$.
\end{theorem}

Define $X_i$ to be $1 - p$ if the expert flips heads (which happens with probability $p$) and $-p$ if the expert flips tails (which happens with probability $1 - p$). Let $Y = \sum_i X_i$. Then $\sigma = \sqrt{p(1 - p)}$ and $\rho = p(1 - p)(p^2 + (1 - p)^2) \le p(1 - p)$. Plugging in these $X_i$ and $n = n_0$ into the Berry-Esseen theorem, we have
\[\abs{F(x) - \Phi(x)} \le \frac{C p(1 - p)}{(p(1 - p))^{3/2} \sqrt{n_0}} = \frac{C}{\sqrt{p(1 - p)n_0}}.\]
Now, we want to approximate $\EE{d_1^\ell} = \EE{\abs{Q(n_0) - p}^\ell}$. Note that $Q(n_0)$ is within $\frac{1}{n_0}$ of $Y + p$, the number of heads flipped divided by $n_0$. This means that $Y - \frac{1}{n_0} \le Q(n_0) - p \le Y + \frac{1}{n_0}$. For this reason, we focus on computing $\EE{\abs{Y}^\ell}$ and subsequently correct for this small difference.

Observe that
\begin{align*}
\EE{\abs{Y}^\ell} &= \int_0^\infty \pr{\abs{Y}^\ell > x} dx = \int_0^\infty \pr{\abs{Y}^\ell > u^\ell} \cdot \ell u^{\ell - 1} du\\
&= \int_0^\infty (\pr{Y > u} + \pr{Y < -u}) \cdot \ell u^{\ell - 1} du = \int_0^\infty\\
&= \int_0^\infty \parens{\pr{\frac{Y\sqrt{n_0}}{\sigma} > \frac{u\sqrt{n_0}}{\sigma}} + \pr{\frac{Y\sqrt{n_0}}{\sigma} < \frac{-u\sqrt{n_0}}{\sigma}}} \cdot \ell u^{\ell - 1} du\\
&= \int_0^\infty \parens{1 - F \parens{\frac{u \sqrt{n_0}}{\sigma}} + F \parens{\frac{-u \sqrt{n_0}}{\sigma}}} \cdot \ell u^{\ell - 1} du.
\end{align*}
Now, observe on the other hand that
\[\int_0^\infty \parens{1 - \Phi \parens{\frac{u \sqrt{n_0}}{\sigma}} + \Phi \parens{\frac{-u \sqrt{n_0}}{\sigma}}} \cdot \ell u^{\ell - 1} du = \EE{\abs{Z}^\ell},\] 
where $Z$ is a random variable drawn from a normal distribution with mean zero and variance $\frac{\sigma^2}{n_0}$. Now we ask: how different is this second quantity (the integral involving $\Phi$) from the first one (the integral involving $F$)?

The answer is, not that different. Indeed, as we derived, $F \parens{\frac{u \sqrt{n_0}}{\sigma}}$ is within $\frac{C}{\sqrt{p(1 - p)n_0}}$ of $\Phi \parens{\frac{u \sqrt{n_0}}{\sigma}}$ for all $u$. Furthermore, for any $r < \frac{1}{2}$, if $u \le -\frac{\sqrt{p(1 - p)}}{n_0^r}$ then $F \parens{\frac{u\sqrt{n_0}}{\sigma}}$ and $\Phi \parens{\frac{u\sqrt{n_0}}{\sigma}}$ are both $e^{-n_0^{\Omega(1)}}$. (For $\Phi$ this follows by concentration of normal distributions; the claim for $F$ follows from Claim~\ref{claim:omega_unlikely}, realizing the fact that there is nothing special about the $0.49$ in the exponent except that it is less than $\frac{1}{2}$.) Similarly, if $u \ge \frac{\sqrt{p(1 - p)}}{n_0^r}$ then both $F \parens{\frac{u\sqrt{n_0}}{\sigma}}$ and $\Phi \parens{\frac{u\sqrt{n_0}}{\sigma}}$ are exponentially close to $1$. Finally, for $u \le -1$ we have $F \parens{\frac{u\sqrt{n_0}}{\sigma}} = 0$ and $\Phi \parens{\frac{u\sqrt{n_0}}{\sigma}} = O(e^{-u^2 n_0})$. This means that $\EE{\abs{Y}^\ell}$ is within
\[\int_0^{\frac{\sqrt{p(1 - p)}}{n_0^{r}}} \frac{2C}{\sqrt{p(1 - p)n_0}} \cdot \ell u^{\ell - 1} du + e^{-n_0^{\Omega(1)}} = 2C(p(1 - p))^{(\ell - 1)/2} n_0^{-r\ell - 1/2} + e^{-n_0^{\Omega(1)}}\]
of $\EE{\abs{Z}^\ell}$. 
Now, since $p(1 - p) \ge c^t \ge c^{.3}$ and $n_0 \ge \frac{1}{\alpha c^{1/3}}$, we have that $p(1 - p) \ge (\alpha n_0)^{-.9}$. It is easy to check that setting any $r > \frac{1}{2} - \frac{1}{20\ell}$ shows that
\[\EE{\abs{Y}^\ell} = \EE{\abs{Z}^\ell}(1 + o(1))\]
where the $o(1)$ depends only on $c$, not on $p$. Note that $\EE{\abs{Z}^\ell} = \Theta \parens{\frac{p(1 - p)}{n_0}}^{\ell/2} = \omega \parens{\frac{1}{n_0^\ell}}$ uniformly over $p$. This means that $\frac{1}{n_0^\ell} = o \parens{\EE{\abs{Y}}^\ell}$ uniformly over $p$.

Now, recall that $Y - \frac{1}{n_0} \le Q(n_0) - p \le Y + \frac{1}{n_0}$, so
\[\max \parens{0, \abs{Y} - \frac{1}{n_0}} \le \abs{Q(n_0) - p} \le \abs{Y} + \frac{1}{n_0},\]
and thus
\[\EE{\max \parens{0, \abs{Y} - \frac{1}{n_0}}^\ell} \le \EE{\abs{Q(n_0) - p}^\ell} = \EE{d_1} \le \EE{\parens{\abs{Y} + \frac{1}{n_0}}^\ell}.\]
By the first statement of Proposition~\ref{prop:exp_l} (with $X_{c,p} = \abs{Y}$ and $Y_{c,p} = \frac{1}{n_0}$), we have that
\[\lim_{c \to 0} \frac{\EE{\parens{\abs{Y} + \frac{1}{n_0}}^\ell}}{\EE{\abs{Y}^\ell}} = 1\]
uniformly over $p$. By the second statement (with $X_{c,p} = \max \parens{0, \abs{Y} - \frac{1}{n_0}}$ and $Y_{c,p} = \abs{Y} - \max \parens{0, \abs{Y} - \frac{1}{n_0}}$), we have that
\[\lim_{c \to 0} \frac{\EE{\max \parens{0, \abs{Y} - \frac{1}{n_0}}^\ell}}{\EE{\abs{Y}^\ell}} = 1\]
uniformly over $p$. (The premise of the second statement is satisfied because $\abs{Y} - \max \parens{0, \abs{Y} - \frac{1}{n_0}} \le \frac{1}{n_0}$.) Therefore, the squeeze theorem tells us that
\[\lim_{c \to 0} \frac{\EE{d_1}}{\EE{\abs{Y}^\ell}} = 1\]
uniformly over $p$. Therefore, we have $\EE{d_1} = \EE{\abs{Z}^\ell}(1 + o(1))$ where the $o(1)$ term only depends on $c$.

Finally, note that
\[\EE{d_1} = \EE{d_1 \mid \Omega_N} \pr{\Omega_N} + \EE{d_1 \mid \overline{\Omega_N}} \pr{\overline{\Omega_N}}.\]
Since $0 \le \EE{d_1 \mid \overline{\Omega_N}} \le 1$ and $\pr{\overline{\Omega_N}} = e^{-n_0^{\Omega(1)}}$ (where the $\Omega(1)$ does not depend on $p$), we have that $\EE{d_1 \mid \Omega_N}$ is within $e^{-n_0^{\Omega(1)}}$ of $\EE{d_1}$. Applying Proposition~\ref{prop:exp_l} in the same way as earlier, we find that $\EE{d_1^\ell \mid \Omega_N} = \EE{d_1^\ell}(1 + o(1)) = \EE{\abs{Z}^\ell}(1 + o(1))$. We know that $\EE{\abs{Z}^\ell} = \mu_\ell \parens{\frac{p(1 - p)}{n_0}}^{\ell/2}$. This completes the proof.
\end{proof}

We can now prove Claim~\ref{claim:d2_o_d1}.

\begin{proof}[Proof (of Claim~\ref{claim:d2_o_d1})]
Let $k = n_{\text{stop}} - n_0$. Define $\{Y_i\}_{i = 0}^k$ as follows: $Y_0 = 0$ and for $i > 0$, $Y_i$ is either $Y_{i - 1} + 1 - p$ (if the $n_0 + i$-th flip is heads, i.e. with probability $p$) or $Y_{i - 1} - p$ (if the $n_0 + i$-th flip is tails, i.e. with probability $1 - p$. Note that $\{Y_i\}$ is a martingale.

Now, observe that for any $0 \le i \le k$, we have
\[Q(n_0 + i) = \frac{(n_0 + 2)Q(n_0) + Y_i + pi}{n_0 + i + 2}.\]
This is because $(n_0 + 2)Q(n_0)$ is one more than the number of heads in the first $n_0$ flips and $(n_0 + i + 2)Q(n_0 + i)$ is one more than the number of heads in the first $n_0 + i$ flips. Thus,
\[\abs{Q(n_0 + i) - Q(n_0)} = \abs{\frac{Y_i + i(p - Q(n_0))}{n_0 + i + 2}} \le \frac{\max_i \abs{Y_i} + k \abs{p - Q(n_0)}}{n_0} = \frac{\max_i \abs{Y_i} + kd_1}{n_0}.\]
Therefore we have
\begin{align*}
\EE{d_2^\ell} &\le \EE{\max_{i \le k} \parens{\abs{Q(n_0 + i) - Q(n_0)}^\ell}} \le \frac{\EE{(\max_i \abs{Y_i} + kd_1)^\ell}}{n_0^\ell} = \frac{2^\ell \EE{\parens{\frac{\max_i \abs{Y_i} + kd_1}{2}}^\ell}}{n_0^\ell}\\
&\le \frac{2^\ell \EE{\frac{(\max_i \abs{Y_i})^\ell + (kd_1)^\ell}{2}}}{n_0^\ell} = \frac{2^{\ell - 1}}{n_0^\ell} \parens{\EE{\parens{\max_i \abs{Y_i}}^\ell} + k^\ell \EE{d_1^\ell}}.
\end{align*}
Here, the last inequality follows from the fact that the arithmetic mean of $\max_i \abs{Y_i}$ and $kd_1$ is less than or equal to the $\ell$-power mean (since $\ell \ge 1$).

Now, it is clear that $\frac{2^{\ell - 1}k^\ell}{n^\ell} \EE{d_1^\ell} = o(\EE{d_1^\ell})$, since $k = o(n_0)$ by Proposition~\ref{prop:n_stop_bound}. We now show that $\frac{2^{\ell - 1}}{n_0^\ell}\EE{\parens{\max_i \abs{Y_i}}^\ell} = o(\EE{d_1^\ell})$. We make use of a tool called the Burkholder-Davis-Gundy inequality.

\begin{defin}
Let $Y = \{Y_i\}_{i = 0}^k$ be a martingale. The \emph{quadratic variation} of $Y$, denoted $[Y]$, is equal to
\[[Y] = \sum_{i = 1}^k (Y_i - Y_{i - 1})^2.\]
Note that $[Y]$ is a random variable, not a number.
\end{defin}
\begin{theorem}[Burkholder-Davis-Gundy inequality]
Let $\ell \ge 1$. There is a constant $C_\ell$ such that for every martingale $Y = \{Y_i\}_{i = 0}^k$ with $Y_0 = 0$, we have
\[\EE{\parens{\max_{i = 0}^k \abs{Y_i}}^\ell} \le C_\ell \EE{[Y]^{\ell/2}}.\]
\end{theorem}

We wish to bound $\EE{\parens{\max_{i = 0}^k \abs{Y_i}}^\ell}$ above. To do so, we bound $\EE{[Y]^{\ell/2}}$ above. Observe that $[Y]$ is a sum of $k$ independent random variables that are each either $p^2$ (with probability $1 - p$) or $(1 - p)^2$ (with probability $p$). Thus, $\EE{[Y]} = kp(1 - p) \equiv \mu$. Observe that
\begin{align*}
\EE{[Y]^{\ell/2}} &= \int_0^\infty \pr{[Y]^{\ell/2} \ge x} dx = \int_0^\infty \pr{[Y] \ge x^{2/\ell}} dx \le (3\mu)^{\ell/2} + \int_{(3\mu)^{\ell/2}}^\infty \pr{[Y] \ge x^{2/\ell}} dx\\
&\le (3\mu)^{\ell/2} + \int_{(3\mu)^{\ell/2}}^\infty e^{\frac{\mu - x^{2/\ell}}{2}} dx.
\end{align*}
The last line comes from a Chernoff bound. In particular, we have that $\pr{[Y] \ge \mu(1 + \delta)} \le e^{-\delta^2 \mu/(2 + \delta)} \le e^{-\delta \mu/2}$ for $\delta \ge 2$. Setting $\delta = \frac{x^{2/\ell}}{\mu} - 1$ gives us the expression above. Now, we can bound the integral as follows:
\[\int_{(3\mu)^{\ell/2}}^\infty e^{\frac{\mu - x^{2/\ell}}{2}} dx \le ((5\mu)^{\ell/2} - (3\mu)^{\ell/2}) e^{-\mu} + ((7\mu)^{\ell/2} - (5\mu)^{\ell/2}) e^{-2\mu} + \dots \equiv B(\mu).\]
Note that $B(\mu)$ is continuous, converges on $[0, \infty)$, and approaches zero as $\mu \to \infty$. It follows that $B(\mu)$ is bounded on $[0, \infty)$; in other words, our integral is $O(1)$ (i.e. possibly depends on $\ell$ but is at most a constant for fixed $\ell$). Therefore, we have
\[\EE{\parens{\max_i \abs{Y_i}}^\ell} \le \EE{[Y]^{\ell/2}} \le 3^{\ell/2}(kp(1 - p))^{\ell/2} + O(1).\]
Therefore we have
\[\frac{2^{\ell - 1}}{n_0^\ell}\EE{\parens{\max_i \abs{Y_i}}^\ell} = O \parens{\parens{\frac{p(1 - p)}{n_0}}^{\ell/2} \parens{\frac{k}{n_0}}^{\ell/2}} = O \parens{\parens{\frac{k}{n_0}}^{\ell/2} \EE{d_1^\ell}} = o(\EE{d_1^\ell}).\]
Therefore, we have that $\EE{d_2^\ell} = o(\EE{d_1^\ell})$. By the same reasoning as in the proof of Claim~\ref{claim:be}, it follows that $\EE{d_2 \mid \Omega_N} = o(\EE{d_1^\ell})$. We previously showed that $\EE{d_1^\ell} = \Theta(\EE{d_1^\ell \mid \Omega_N})$. This completes the proof.
\end{proof}

\begin{theorem} \label{thm:local}
If $f$ is a respectful, normalizable, continuously differentiable proper scoring rule, and $\err_c(p)$ is the expected error of a locally adaptive expert rewarded by $f$ when the coin has bias $p$ and the cost of a flip is $c$, then
\[\lim_{c \to 0} c^{-\ell/4} \EE[p \leftarrow U_{[0, 1]}]{\err_c(p)^\ell} = \mu_\ell \int_0^1 \parens{\frac{2x(1 - x)}{R''(x)}}^{\ell/4} dx.\]
\end{theorem}

\begin{proof}[Proof (of Theorem~\ref{thm:local})]
Let $N = \frac{1}{\alpha c^{1/3}}$, i.e. a large enough function of $c$ that it is guaranteed that the expert flips the coin at least $N$ times. We have
\[c^{-\ell/4} \EE[p \leftarrow U_{[0, 1]}]{\err_c(p)^\ell} = c^{-\ell/4}\parens{\EE[p \leftarrow U_{[0, 1]}]{\err_c(p)^\ell \mid \Omega_N} \pr{\Omega_N} + \EE[p \leftarrow U_{[0, 1]}]{\err_c(p)^\ell \mid \overline{\Omega_N}} \pr{\overline{\Omega_N}}}.\]
We wish to compute the limit of this quantity as $c$ approaches zero. Note that
\[\lim_{c \to 0} c^{-\ell/4} \EE[p \leftarrow U_{[0, 1]}]{\err_c(p)^\ell \mid \overline{\Omega_N}} \pr{\overline{\Omega_N}} = 0.\]
This is because $\err_c(p)$ is bounded between $0$ and $1$ and $\pr{\overline{\Omega_N}} = O(e^{-N^{.01}}) = O(e^{-\Omega(c^{-1/300})})$, which goes to zero faster than $c^{-\ell/4}$ goes to infinity. Therefore we have
\[\lim_{c \to 0} c^{-\ell/4} \EE[p \leftarrow U_{[0, 1]}]{\err_c(p)^\ell} = \lim_{c \to 0} c^{-\ell/4} \EE[p \leftarrow U_{[0, 1]}]{\err_c(p)^\ell \mid \Omega_N}.\]
(We may ignore the $\pr{\Omega_N}$ term above because it approaches $1$ in the limit.) We may write this quantity as
\[\lim_{c \to 0} c^{-\ell/4} \parens{(1 - 4c^t) \EE[p \leftarrow U_{[2c^t, 1 - 2c^t]}]{\err_c(p)^\ell \mid \Omega_N} + 4c^t \EE[p \leftarrow U_{[0, 2c^t] \cup [1 - 2c^t, 1]}]{\err_c(p)^\ell \mid \Omega_N}}.\]
Let us focus on the second summand. Let $p \in [0, 2c^t] \cup [1 - 2c^t, 1]$. We assume $p \in [0, 2c^t]$; the other case is analogous.

We consider two sub-cases: $p \in [0, \alpha c^{1/3}]$ and $p \in [\alpha c^{1/3}, 2c^t]$. First suppose that $p \in [c^{1/3}, 2c^t]$. Note that since $\Omega_N$ holds, we have for all $n \ge N$ that
\[\abs{Q(n) - p} \le \frac{\sqrt{p(1 - p)}}{n^{.49}} \le \frac{\sqrt{2c^t}}{N^{.49}} = \sqrt{2}\alpha^{.49} c^{t/2 + .49/3}.\]
This in particular is true of $n = \nstop$, so
\[\err_c(p)^\ell \le (\sqrt{2}\alpha^{.49})^\ell c^{\ell(t/2 + .49/3)} = o(c^{\ell/4})\]
since $t > \frac{1}{4}$ and so $\frac{t}{2} + \frac{.49}{3} > \frac{1}{4}$.

Now suppose that $p \in [0, \alpha c^{1/3}]$. Recall the notation $Q_p(n)$ from the discussion preceding the definition of $\Omega_N$. For any $n \ge N$, we have
\begin{align*}
\abs{Q(n) - p} &\le \abs{Q(n) - Q_{\alpha c^{1/3}}(n)} + \abs{Q_{\alpha c^{1/3}}(n) - \alpha c^{1/3}} + \abs{\alpha c^{1/3} - p}\\
&\le Q_{\alpha c^{1/3}}(n) + \abs{Q_{\alpha c^{1/3}}(n) - \alpha c^{1/3}} + \alpha c^{1/3} \le 2\alpha c^{1/3} + 2\abs{Q_{\alpha c^{1/3}}(n) - \alpha c^{1/3}}\\
&\le 2\alpha c^{1/3} + \frac{2\sqrt{\alpha c^{1/3}}}{n^{.49}} \le 2\alpha c^{1/3} + 2\sqrt{\alpha} c^{1/6 + .49/3} = o(c^{1/4})
\end{align*}
so $\err_c(p)^\ell = o(c^{\ell/4})$.

This means that
\[\lim_{c \to 0} c^{-\ell/4} \cdot 4c^t \EE[p \leftarrow U_{[0, 2c^t] \cup [1 - 2c^t, 1]}]{\err_c(p)^\ell \mid \Omega_N} = 0\]
so we can ignore this summand. Therefore, we have
\[\lim_{c \to 0} c^{-\ell/4} \EE[p \leftarrow U_{[0, 1]}]{\err_c(p)^\ell} = \lim_{c \to 0} c^{-\ell/4} (1 - 4c^t) \EE[p \leftarrow U_{[2c^t, 1 - 2c^t]}]{\err_c(p)^\ell \mid \Omega_N}.\]
From Lemma~\ref{lem:close}, we have that
\begin{align*}
(1 - o(1)) \int_{2c^t}^{1 - 2c^t} \mu_\ell \parens{\frac{2x(1 - x)}{R''(x)}}^{\ell/4} dx &\le c^{-\ell/4} (1 - 4c^t) \EE[p \leftarrow U_{[2c^t, 1 - 2c^t]}]{\err_c(p)^\ell \mid \Omega_N}\\
&\le (1 + o(1)) \int_{2c^t}^{1 - 2c^t} \mu_\ell \parens{\frac{2x(1 - x)}{R''(x)}}^{\ell/4} dx.
\end{align*}
By the squeeze theorem, we conclude that.
\[\lim_{c \to 0} c^{-\ell/4} \EE[p \leftarrow U_{[0, 1]}]{\err_c(p)^\ell} = \lim_{c \to 0} \int_{2c^t}^{1 - 2c^t} \mu_\ell \parens{\frac{2x(1 - x)}{R''(x)}}^{\ell/4} dx = \mu_\ell \int_0^1 \parens{\frac{2x(1 - x)}{R''(x)}}^{\ell/4} dx.\]
\end{proof}

Below is our proof of Lemma~\ref{lem:global_helper}, followed by a proof of Theorem~\ref{thm:global}. Our approach will be to compare the behavior of a locally adaptive expert to that of a globally adaptive one. We will assume that the experts observe the same stream of coin flips (each heads with probability $p$ unknown to the experts) but that they may decide to stop at different times. As before, we will let $Q(n) = \frac{h + 1}{n + 2}$ where $h$ is the number of the first $n$ flips to have come up heads; since the experts see the same coin flips, we do not need to distinguish between $Q(n)$ for the locally adaptive expert and for the globally adaptive expert. We will let $n_l$ and $n_g$ be the number of times the locally and globally adaptive experts flip the coin, respectively (so $n_g \ge n_l$). (We used the notation $\nstop$ in place of $n_l$ in Proposition~\ref{prop:n_stop_bound}.) Let $t$ be as in the definition of respectful scoring rules, and in particular we will assume that $t < 0.3$ as before (for any $t$ that witnesses that a scoring rule is respectful, any smaller $t > \frac{1}{4}$ also works).

The bulk of the proof of Theorem~\ref{thm:global} has already been completed, if we think of Theorem~\ref{thm:local} as a step in the proof. The bulk of the remainder is proving the following lemma.

\begin{proof}[Proof of Lemma~\ref{lem:global_helper}]
Suppose the globally adaptive expert flips the coin $N := (1 + 6\gamma c^{1/300}) n_l$ times. We show that they do not flip the coin another time.

By definition of $\Omega_{n_l}$, we have that
\[p - \frac{\sqrt{p(1 - p)}}{n_l^{.49}} \le Q(n_l) \le p + \frac{\sqrt{p(1 - p)}}{n_l^{.49}}.\]
It is easy to check that because $n_l = \Omega(c^{-1/3})$ (by Claim~\ref{claim:one_third_bound}) and $4c^t \le Q(n_l) \le 1 - 4c^t$ (so $Q(n_l), 1 - Q(n_l) = \Omega(n_l^{-.9})$), the above relationship between $Q(n_l)$ and $p$ implies that $2c^t \le p \le 1 - 2c^t$. This allows us to use some results from our analysis of locally adaptive experts. In particular, by Equation~\ref{eq:delta_bound} in the proof of Proposition~\ref{prop:n_stop_bound}, we have that
\[\frac{p(1 - p)}{2n_l^2}R''(p)(1 - \gamma c^{1/300}) \le \Delta_{n_l + 1} \le \frac{p(1 - p)}{2n_l^2}R''(p)(1 + \gamma c^{1/300}).\]
Conditional on $\Omega_N$ (and by definition $\Omega_{n_l}$ implies $\Omega_N$), we also have
\[\frac{p(1 - p)}{2N^2}R''(p)(1 - \gamma c^{1/300}) \le \Delta_{N + 1} \le \frac{p(1 - p)}{2N^2}R''(p)(1 + \gamma c^{1/300}).\]
In particular this means that
\[\Delta_{N + 1} \le \frac{n_l^2}{N^2} \cdot \frac{1 + \gamma c^{1/300}}{1 - \gamma c^{1/300}} \Delta_{n_l + 1} \le \parens{\frac{n_l}{N}}^2(1 + 3\gamma c^{1/300}) \Delta_{n_l + 1}\]
for $c$ small enough. This means that if $\parens{\frac{n_l}{N}}^2 \le \frac{1}{1 + 6\gamma c^{1/300}}$ then $\Delta_{N + 1} \le (1 - 2\gamma c^{1/300}) \Delta_{n_l + 1} < c(1 - 2\gamma c^{1/300})$. Furthermore, for any $n \ge N$ we will have $\Delta_{n + 1} < c(1 - 2\gamma c^{1/300})$.

However, this does not mean that the globally adaptive expert won't flip the coin for the $N + 1$-th time, because they don't know that $\Omega_{n_l}$ is true. From the expert's perspective, \emph{if} they knew $\Omega_{n_l}$ (or even $\Omega_N$) to be true, they would stop flipping the coin; but perhaps they should keep flipping the coin because of the outside chance that $\Omega_N$ is false.

This turns out not to be the case, because the probability that $\Omega_N$ is false is so small. In particular, from the expert's perspective, if $\Omega_N$ being false, they cannot achieve reward better than the expectation of $R(p)$ conditional on the coins they've flipped and on $\Omega_N$ being false. We show that if the scoring rule $f$ is normalizable (i.e. $\int_0^1 R(x) dx$ is finite), then this quantity isn't too large. In particular, we show the following:

\begin{claim}
Let $H_N$ be the random variable corresponding to the number of heads flipped in the first $N$ flips. Then for any $3c^tN \le h \le (1 - 3c^t)N$, we have
\[\EE{R(p) \mid \overline{\Omega_N}, H_N = h} \le \frac{2}{c^t} \int_0^1 R(x) dx.\]
\end{claim}

\begin{proof}
We have
\[\EE{R(p) \mid \overline{\Omega_N}, H_N = h} \le \EE{R(p) \mid \overline{\Omega_N}, H_N = h, p < \frac{1}{2}} + \EE{R(p) \mid \overline{\Omega_N}, H_N = h, p > \frac{1}{2}}.\]
Let us consider the expectation conditioned on $p < \frac{1}{2}$. Consider the distribution $D$ of $p$ conditioned on $\overline{\Omega_N}$, $H_N = h$, and $p < \frac{1}{2}$. Consider also the uniform distribution $D'$ on $[0, c^t]$.

We claim that $D$ stochastically dominates $D'$, i.e. $\pr[x \leftarrow D]{x \le y} \le \pr[x \leftarrow D']{x \le y}$ for all $y$. To see this, observe that the PDF of $D$ is an increasing function on $[0, c^t]$. This is because $D$ on $[0, c^t]$ is a constant multiple of the distribution $D''$ of $p$ conditioned on $\overline{\Omega_N}$, $H_N = h$, and $p \le c^t$; but in this case the condition $\overline{\Omega_N}$ is redundant because if $p \le c^t$ then $\overline{\Omega_N}$ holds. So $D''$ is the distribution of $p$ conditioned on $H_N = h$ and $p \le c^t$. Clearly the PDF of $D''$ increases on $[0, c^t]$ (because the expert starts with uniform priors and updates more strongly on against values of $p$ farther from $\frac{h + 1}{N + 2}$, which is greater than $2.9c^t$ for $c$ small enough).

Now, the expectation of $R(p)$ if $p$ were drawn from $D'$ instead of $D$ is equal to $\frac{1}{c^t} \int_0^{c^t} R(x) dx \le \frac{1}{c^t} \int_0^1 R(x) dx$. On the other hand, the actual expectation of $R(p)$ (i.e. with $p$ drawn from $D$) is necessarily smaller. This is because $R$ is convex and symmetric about $\frac{1}{2}$, meaning that $R$ is decreasing on $(0, \frac{1}{2})$. Since $D$ stochastically dominates $D'$, we conclude that
\[\EE{R(p) \mid \overline{\Omega_N}, H_N = h, p < \frac{1}{2}} \le \frac{1}{c^t} \int_0^1 R(x) dx.\]
The same inequality holds conditional instead on $p > \frac{1}{2}$, which concludes the proof.
\end{proof}

From the expert's perspective, this means that if they flip the coin for the $N + 1$-th time, then:
\begin{itemize}
\item In the case that $\Omega_N$ is true, the best case is that they never flip the coin again, in which case they will pay a total cost of $c$ and get expected reward at most $c(1 - 2\gamma c^{1/300})$.
\item In the case that $\Omega_N$ is false, the best case is that they get reward $\frac{2}{c^t} \int_0^1 R(x) dx$.
\end{itemize}
In other words, the expert's expected reward if they flip the coin for the $N + 1$-th time and pursue the optimal strategy from there is at most
\[\frac{2}{c^t} \int_0^1 R(x) dx \cdot \pr{\overline{\Omega_N}} - 2\gamma c^{1/300} \pr{\Omega_N} \le O \parens{\frac{e^{-N^{.01}}}{c^t}} - \gamma c^{1/300} = O \parens{\frac{e^{-\Omega(c^{1/300})}}{c^t}} - \gamma c^{1/300}.\]
The first step is nontrivial: it uses the fact that the probability that the expert assigns to $\overline{\Omega_N}$ after the first $N$ flips is $O(e^{-N^{.01}})$. This doesn't immediately follow from Claim~\ref{claim:omega_unlikely} because the claim only states that the prior probability of $\overline{\Omega_N}$, i.e. before any flips, is $O(e^{-N^{.01}})$. To see that the posterior probability (after the first $N$ flips) is also of this order, we first observe that the posterior probability cannot depend on the order of the flip outcomes; this is apparent from the definition of $\Omega_N$. However, perhaps the \emph{number} of heads, i.e. the value of $H_N$, affects the posterior probability of $\overline{\Omega_N}$. This may be so, but it cannot increase the probability by more than a factor of $N + 1$. That is because the prior for $H_N$ is uniform over $\{0, \dots, N\}$.\footnote{For any $h \in \{0, \dots, N\}$, the prior probability that $H_N = h$ is given by $\int_{p = 0}^1 \binom{N}{h} p^h (1 - p)^{N - h} dp$, and this integral evaluates to $\frac{1}{N + 1}$ (see Fact~\ref{fact:beta}).}

Now, the quantity on the right is negative for $c$ small enough, so the expert will not flip the $N + 1$-th coin. This proves the claim.
\end{proof}

The following corollary is essentially identical to Proposition~\ref{prop:n_stop_bound} but for globally adaptive experts.
\begin{corollary} \label{cor:global_close}
Assume that $\Omega_N$ holds for some $N$. For sufficiently small $c$, for all $p \in [8c^t, 1 - 8c^t]$, we have
\[\sqrt{\frac{p(1 - p)R''(p)}{2c}(1 - 14\gamma c^{1/300})} \le n_g \le \sqrt{\frac{p(1 - p)R''(p)}{2c}(1 + 14\gamma c^{1/300})}.\]
\end{corollary}

\begin{proof}
Because $\Omega_N$ holds for some $N$, for sufficiently small $c$ the fact that $p \in [8c^t, 1 - 8c^t]$ implies that $Q(n_l) \in [4c^t, 1 - 4c^t]$. This means that we may apply Lemma~\ref{lem:global_helper} to say that $n_g \le N$. Consequently we have that
\begin{align*}
&\sqrt{\frac{p(1 - p)R''(p)}{2c}(1 - \gamma c^{1/300})} \le n_l \le n_g \le N = (1 + 6\gamma c^{1/300}) n_l\\
&\le (1 + 6\gamma c^{1/300}) \sqrt{\frac{p(1 - p)R''(p)}{2c}(1 + \gamma c^{1/300})} \le \sqrt{\frac{p(1 - p)R''(p)}{2c}(1 + 14\gamma c^{1/300})}
\end{align*}
for $c$ small enough. Therefore we have
\[\sqrt{\frac{p(1 - p)R''(p)}{2c}(1 - 14\gamma c^{1/300})} \le n_g \le \sqrt{\frac{p(1 - p)R''(p)}{2c}(1 + 14\gamma c^{1/300})}.\]
\end{proof}

Theorem~\ref{thm:global} follows as a simple corollary.

\begin{proof}[Proof (of Theorem~\ref{thm:global})]
The lemma analogous to Lemma~\ref{lem:close} but for expected globally adaptive error, and for $p \in [8c^t, 1 - 8c^t]$, follows immediately from Corollary~\ref{cor:global_close}. This is because the proof of Lemma~\ref{lem:close} makes no assumptions about the specific value of $\gamma$ (other than that it is positive), which means that the proof goes through just as well for $14\gamma$ in place of $\gamma$. Theorem~\ref{thm:global} follows from this fact exactly in the same way that Theorem~\ref{thm:local} followed from Lemma~\ref{lem:close}.
\end{proof}
\section{Omitted Proofs from Section~\ref{sec:optimal}}\label{app:optimal}
\subsection{Proof Overview for Theorem~\ref{thm:optimal}}
As shown in Corollary~\ref{cor:strictlyproper}, the equation $xg'(x) = (1 - x)g'(1 - x)$ lets us extend $g$ uniquely in a continuous manner to $(0, 1)$ if we know $g$ on $[\half, 1)$. Thus, we can simply consider $g$ on $[\half, 1)$. For $g'$ to be nonnegative everywhere, it suffices for it to be nonnegative on $[\half, 1)$, because of the relation $xg'(x) = (1 - x)g'(1 - x)$. Also, observe that the integrand is symmetric about $\half$; this is clear from the fact that $R''$ is symmetric about $\half$. This means that
\[\ind^\ell(g) = 2 \int_{\half}^1 \parens{\frac{x(1 - x)^2}{g'(x)}}^{\ell/4} dx.\]
Thus, our question can be phrased as follows: find the continuously differentiable function $g: [\half, 1) \to \RR$ satisfying $g \parens{\half} = 0$, $g'(x) \ge 0$, and $\int_\half^1 (1 - x)g'(x) dx = 1$, that minimizes
\[\int_{\half}^1 \parens{\frac{x(1 - x)^2}{g'(x)}}^{\ell/4} dx.\]

From this point, our problem is simply a continuous mathematical program. It is not obvious that the program should admit a closed-form solution, but it does. We defer all details to Appendix~\ref{app:optimal}, and just briefly note that we can formulate the problem exclusively as a function of $g'$, and then uniquely reconstruct $g$ using $g(\half) = 0$. Once we have done this, we can take a Lagrangian relaxation by putting a multiplier on the constraint $\int_\half^1 (1 - x)g'(x) dx = 1$, and hope that the solution to the relaxation is continuous and satisfies $g'(x) \geq 0$. While this is not guaranteed to succeed, this method does in fact nail down the optimum. The main technical lemmas which yield Theorem~\ref{thm:optimal} are:

\begin{lemma} \label{lem:besth}
For any $\ell \ge 1$, a function $h: [\half, 1) \to \RR_{\ge 0}$ satisfying $\int_\half^1 (1 - x)h(x) dx = 1$ that minimizes $\int_{\half}^1 \parens{\frac{x(1 - x)^2}{g'(x)}}^{\ell/4} dx$ is $\tilde{h}_\ell(x) = \kappa_\ell (x^\ell (1 - x)^{2\ell - 4})^{1/(\ell + 4)}$, where $\kappa_\ell = \parens{\int_\half^1 (x(1 - x)^3)^{\ell/(\ell + 4)} dx}^{-1}$.
\end{lemma}

(Note that $\kappa_\ell$ is simply a normalization constant, so as to make $\int_\half^1 (1 - x)h(x) dx$ equal $1$.)

\begin{corollary} \label{cor:hunique}
The unique continuous function $h: [\half, 1) \to \RR_{\ge 0}$ satisfying $\int_\half^1 (1 - x)h(x) dx = 1$ that minimizes $\int_{\half}^1 \parens{\frac{x(1 - x)^2}{h(x)}}^{\ell/4} dx$ is $\tilde{h}_\ell$.
\end{corollary}

Theorem~\ref{thm:optimal} then follows from Corollary~\ref{cor:hunique} by setting $g_{\ell,\opt}$ to the integral of $\tilde{h}_\ell$ on $[1/2,1)$, and extending it to $(0,1/2)$ via Corollary~\ref{cor:strictlyproper}. For some choices of $\ell$, the particular scoring rule $g_{\ell,\opt}$ has an interesting closed form (see Section~\ref{sec:compare} below), but this is not true for all $\ell$. Even in cases where the particular closed form is not illuminating, the fact that $g_{\ell,\opt}$ even exists is already interesting, and the fact that Theorem~\ref{thm:optimal} nails down the closed form allows us to compare other scoring rules to the optimum. We conclude with a remark, confirming that our analysis in Section~\ref{sec:index} indeed is meaningful for all derived optimal scoring rules. A proof for $\ell \in [1,8]$ is in Appendix~\ref{app:index}, and a proof for $\ell > 8$ is in Appendix~\ref{app:weierstrass}.

\begin{remark}\label{rem:respect}
For every $\ell \in [1, 8]$, $g_{\ell, \opt}$ is respectful. For $\ell > 8$, and all $\varepsilon > 0$, there exists a respectful normalized proper scoring rule $g(\cdot)$ such that $\abs{g(x) - g_{\ell, \opt}(x)} \le \varepsilon$ for all $x \in (0, 1)$, with $\ind^\ell(g) \leq \ind^\ell(g_{\ell,\opt})+\varepsilon$.
\end{remark}

A corollary of Remark~\ref{rem:respect} is the following.
\begin{corollary}
For $\ell \ge 1$, let
\[\errf_\opt^\ell := \inf_f \lim_{c \to 0} c^{-\ell/4} \cdot \errf_c^\ell(f)\]
where $f$ ranges over all normalized, respectful, continuously differentiable proper scoring rules. Let
\[\ind_\opt^\ell := \inf_f \ind^\ell(f)\]
where $f$ ranges over all normalized, continuously differentiable proper scoring rules. Then:
\begin{enumerate}[label=(\arabic*)]
\item \label{item:corollary_1} $\errf_\opt^\ell = \mu_\ell\cdot  2^{\ell/4}\cdot \ind_\opt^\ell$.
\item \label{item:corollary_2} For $1 \le \ell \le 8$, the first infimum is uniquely achieved by $f = g_{\ell, \opt}$.
\item \label{item:corollary_3} For $\ell > 8$, no (respectful) function achieves the first infimum, but the infimum is reached in the limit by uniform approximations of $g_{\ell, \opt}$ (which are normalized, respectful, and continuously differentiable).
\end{enumerate}
\end{corollary}

\subsection{Omitted Proofs}

Letting $h = g'$ and noting that we can uniquely reconstruct $g$ from $h$ using $g \parens{\half} = 0$, we find that we are looking for the continuous function $h: [\half, 1) \to \RR_{\ge 0}$ satisfying $\int_\half^1 (1 - x)h(x) dx = 1$ that minimizes
\[\int_{\half}^1 \parens{\frac{x(1 - x)^2}{h(x)}}^{\ell/4} dx.\]

One such function (which will turn out to be the only continuous one) is given as per the following lemma.

\begin{proof}[Proof of Lemma~\ref{lem:besth}]
Let $\lambda_\ell = \frac{\ell}{4 \kappa_\ell^{\ell/4 + 1}}$. Consider the functional
\begin{align*}
\chi(h) &:= \int_{\half}^1 \parens{\frac{x(1 - x)^2}{h(x)}}^{\ell/4} dx + \lambda_\ell \parens{\int_\half^1 (1 - x)h(x) dx - 1}\\
&= \int_\half^1 \parens{\parens{\frac{x(1 - x)^2}{h(x)}}^{\ell/4} + \lambda_\ell (1 - x)h(x)}dx - \lambda_\ell.
\end{align*}
It suffices to show that among all $h$ satisfying $\int_\half^1 (1 - x)h(x) dx = 1$, $\tilde{h}$ minimizes $\chi(h)$. This is because among such $h$, the second summand in the definition of $\chi$ is always zero. In fact, we prove something stronger: $\tilde{h}$ minimizes $\chi$, among all functions from $[\half, 1)$ to $\RR$. To show this, it suffices to show that for every $x \in [\half, 1)$, the value $y$ that minimizes
\[\parens{\frac{x(1 - x)^2}{y}}^{\ell/4} + \lambda_\ell(1 - x)y\]
is $y = \tilde{h}(x)$. The derivative with respect to $y$ of this expression is
\[\frac{-\ell}{4} (x(1 - x)^2)^{\ell/4} y^{-(\ell/4 + 1)} + \lambda_\ell(1 - x),\]
which is an increasing function of $y$ (since $y^{-(\ell/4 + 1)}$ is a decreasing function of $y$ and $\frac{-\ell}{4} (x(1 - x)^2)^{\ell/4}$ is negative). It is equal to $0$ precisely when
\[y = \parens{\frac{\ell}{4\lambda_\ell}}^{4/(\ell+4)} (x^\ell (1 - x)^{2\ell - 4})^{1/(\ell + 4)} = \kappa_\ell (x^\ell (1 - x)^{2\ell - 4})^{1/(\ell + 4)} = \tilde{h}(x).\]
It remains only to note that $\int_\half^1 (1 - x)\tilde{h}(x) dx = 1$, and this follows immediately from the definition of $\tilde{h}$ and $\kappa_\ell$.
\end{proof}

\begin{proof}[Proof of Corollary~\ref{cor:hunique}]
Suppose for contradiction that there is another continuous function $\hat{h}$ satisfying the above properties that achieves the minimum. Then $\chi(\hat{h}) = \chi(\tilde{h})$, with $\chi$ as in the proof of Lemma~\ref{lem:besth}, since $\tilde{h}$ minimizes $\chi$ and the second summand in the definition of $\chi$ is zero for both $\hat{h}$ and $\tilde{h}$. In particular, we have that $\chi(\hat{h}) - \chi(\tilde{h}) = 0$, i.e.
\[\int_\half^1 \parens{\parens{\parens{\frac{x(1 - x)^2}{\hat{h}(x)}}^{\ell/4} + \lambda_\ell (1 - x)\hat{h}(x)} - \parens{\parens{\frac{x(1 - x)^2}{\tilde{h}(x)}}^{\ell/4} + \lambda_\ell (1 - x)\tilde{h}(x)}} dx = 0.\]
Let $\Delta(x)$ be the integrand. Note that $\Delta$ is always nonnegative, and is zero precisely for those values of $x$ where $\hat{h}(x) = \tilde{h}(x)$ (since, as we showed earlier, $y = \tilde{h}(x)$ is the unique value minimizing $\parens{\frac{x(1 - x)^2}{y}}^{\ell/4} + \lambda_\ell (1 - x)y$). Since $\hat{h} \neq \tilde{h}$, $\Delta$ is positive at some $x_0$; say $\Delta(x_0) = y_0$. Also, note that $\Delta$ is continuous because $\hat{h}$ and $\tilde{h}$ are continuous. This means that for some $\delta > 0$, $\abs{\Delta(x) - y_0} < \frac{y_0}{2}$ for all $x$ such that $x_0 \le x \le x_0 + \delta$. But this means that the integral of $\Delta$ on $[x_0, x_0 + \delta]$ is at least $\frac{\delta \cdot y_0}{2} > 0$, so $\int_\half^1 \Delta(x) > 0$, a contradiction. Therefore, $\tilde{h}$ is indeed the unique continuous function satisfying the stated constraints.
\end{proof}

\begin{proof}[Proof of Theorem~\ref{thm:optimal}]
We have reasoned that $g_{\ell,\text{OPT}}$ is the antiderivative of $\tilde{h}$ on $[\half, 1)$, which gives us $g_{\ell,\text{OPT}}$ for $x \ge \half$. For $x < \half$, we have
\[g_{\ell,\text{OPT}}'(x) = \frac{1 - x}{x} g_{\ell,\text{OPT}}'(1 - x) = \kappa_\ell \frac{1 - x}{x} ((1 - x)^\ell x^{2\ell - 4})^{1/(\ell + 4)} = \kappa_\ell (x^{\ell - 8} (1 - x)^{2\ell + 4})^{1/(\ell + 4)},\]
which extends to the stated function $g_{\ell,\text{OPT}}$ by the fundamental theorem of calculus. The constant we want to add (upon taking the antiderivative) is zero so that $g_{\ell,\text{OPT}} \parens{\half} = 0$. We need to check that $g_{\ell,\text{OPT}}$ is continuously differentiable at $\half$, which means checking that $\tilde{h}$ is continuous at $\half$ when extended to $(0, 1)$. This is indeed the case because
\[\lim_{x \to \half^-} g_{\ell, \text{OPT}}'(x) = \lim_{x \to \half^-} \frac{1 - x}{x} g_{\ell, \text{OPT}}'(1 - x) = \lim_{x \to \half^-} g_{\ell, \text{OPT}}'(1 - x) = \lim_{x \to \half^+} g_{\ell, \text{OPT}}'(x).\]
Finally, $g_{\ell,\text{OPT}}$ is the unique continuous normalized minimizer because its derivative is unique, by Corollary~\ref{cor:hunique}. Note that $g_{\ell,\text{OPT}}$ is in fact strictly proper since $\tilde{h}(x)$ is positive on $[\half, 1)$.
\end{proof}

We finish by noting that for $1 \le \ell \le 8$, the incentivization index really is meaningful for these optimal functions when --- that is, that $g_{\ell, \text{OPT}}$ is respectful for each $\ell \in [1, 8]$. (It is evident that $g_{\ell, \text{OPT}}$ is normalizable, it is in fact normalized.)

\begin{proof}[Proof of Remark~\ref{rem:respect}, part one]
This proof handles the case of $\ell \in [1,8]$.
We have
\[R''_{\ell, \text{OPT}}(x) = \frac{g'_{\ell, \text{OPT}}(x)}{1 - x} = \begin{cases}\kappa_\ell (x^{\ell - 8}(1 - x)^\ell)^{1/(\ell + 4)} & x \le \half \\
\kappa_\ell (x^\ell(1 - x)^{\ell - 8})^{1/(\ell + 4)} & x \ge \half.\end{cases}\]
First note that $R_{\ell, \text{OPT}}$ (henceforth we will simply write $R$) is strongly convex. Since $R''$ is symmetric, it suffices to show this on $(0, \half]$. We have $1 - x \ge \half$ on this interval, and $x^{(\ell - 8)/(\ell + 4)}$ is bounded away from zero when $x \le 8$. Next, the fact that $R'''$ is Riemann integral on any closed sub-interval of $(0, 1)$ is evident. Finally, there are constants $k \neq 0$ and $r$ such that $\lim_{x \to 0} x^r R'''(x) = k$: in particular, $r = \frac{12}{\ell + 4}$ and $k = \frac{(\ell - 8)\kappa_\ell}{\ell + 4}$.

(Note that for $\ell > 8$, $g_{\ell, \opt}$ is not respectful, since $\lim_{x \to 0} R''_{\ell, \opt}(x) = 0$.)
\end{proof}
\section{Omitted Proofs from Section~\ref{sec:weierstrass}}\label{app:weierstrass}

\begin{proof}[Proof of Theorem~\ref{thm:analytic}]
Suppose that $f$ is a proper scoring rule. Then $f$ is nonconstant, $f'(x) \ge 0$ everywhere by Lemma~\ref{lem:weaklyproper}, and by the same lemma we have that
\[xf'(x) = (1 - x)f'(1 - x).\]
Taking successive derivatives of both sides, we have
\begin{align*}
f'(x) + xf''(x) &= -f'(1 - x) - (1 - x)f''(1 - x)\\
2f''(x) + xf'''(x) &= 2f''(1 - x) + (1 - x)f'''(1 - x)\\
3f'''(x) + xf^{(4)}(x) &= -3f'''(1 - x) - (1 - x)f^{(4)}(1 - x)
\end{align*}
and so on. Plugging in $x = \half$, we have
\begin{align*}
f' \parens{\half} + \half f'' \parens{\half} &= -f' \parens{\half} - \half f'' \parens{\half}\\
2f'' \parens{\half} + \half f''' \parens{\half} &= 2f'' \parens{\half} + \half f''' \parens{\half}\\
3f''' \parens{\half} + \half f^{(4)} \parens{\half} &= -3f''' \parens{\half} - \half f^{(4)} \parens{\half}
\end{align*}
and so on. These equations alternate between giving us tautologies and simplifying to the following identities:
\[f'' \parens{\half} = -2f' \parens{\half}; \quad f^{(4)} \parens{\half} = -6f''' \parens{\half}; \quad f^{(6)} \parens{\half} = -10f^{(5)} \parens{\half};\]
and so on, the general form of the identities being that for $k$ odd, we have
\[f^{(k + 1)}\parens{\half} = -2k f^{(k)}\parens{\half}.\]
Since $f$ is analytic, we have
\begin{align*}
f(x) &= \sum_{n = 0}^\infty \frac{1}{n!} f^{(n)} \parens{\half} \parens{x - \half}^n\\
&= f \parens{\half} + \sum_{k > 0 \text{ odd}} \frac{1}{(k + 1)!} \parens{(k + 1) f^{(k)}\parens{\half} \parens{x - \half}^k + f^{(k + 1)} \parens{\half} \parens{x - \half}^{k + 1}}.
\end{align*}
Letting $c_k = \frac{1}{(k + 1)!} f^{(k)} \parens{\half}$ for $k = 0, 1, 3, 5, \dots$, we have
\begin{align*}
f(x) &= c_0 + \sum_{k > 0 \text{ odd}} c_k \parens{(k + 1)\parens{x - \half}^k - 2k \parens{x - \half}^{k + 1}}\\
&= c_0 + \sum_{k > 0 \text{ odd}} c_k(2k + 1 - 2kx)\parens{x - \half}^k.
\end{align*}
This proves the forward direction. Conversely, we claim that if $f$ is nonconstant, $f'(x) \ge 0$ everywhere, and $f$ can be written in the stated form for some $c_0, c_1, c_3, \dots$, then $f$ is a proper scoring rule. We only have to verify that $xf'(x) = (1 - x)f'(1 - x)$ everywhere (by Lemma~\ref{lem:weaklyproper} and Lemma~\ref{lem:constant} (stated and proven below)). Taking the derivative of $(1 - x)f'(1 - x)$ term by term, we have
\begin{align*}
(1 - x)f'(1 - x) &= (x - 1)(f(1 - x))' = (x - 1) \frac{d}{dx} \parens{c_0 + \sum_{k > 0 \text{ odd}} c_k(2kx + 1)\parens{\half - x}^k}\\
&= (x - 1) \sum_{k > 0 \text{ odd}} c_k \parens{2k \parens{\half - x}^k - k(2kx  + 1) \parens{\half - x}^{k - 1}}\\
&= \sum_{k > 0 \text{ odd}} (x - 1)kc_k \parens{\half - x}^{k - 1}(1 - 2x - 2kx - 1)\\
&= \sum_{k > 0 \text{ odd}} 2k(k + 1)c_kx(1 - x)\parens{\half - x}^{k - 1}.
\end{align*}

Similarly, we have
\begin{align*}
xf'(x) &= x \frac{d}{dx} \parens{c_0 + \sum_{k > 0 \text{ odd}} c_k(2k + 1 - 2kx)\parens{x - \half}^k}\\
&= x \sum_{k > 0 \text{ odd}} c_k \parens{-2k \parens{x - \half}^k + k(2k + 1 - 2kx) \parens{x - \half}^{k - 1}}\\
&= \sum_{k > 0 \text{ odd}} xkc_k \parens{x - \half}^{k - 1}(1 - 2x + 2k + 1 - 2kx)\\
&= \sum_{k > 0 \text{ odd}} 2k(k + 1)c_kx(1 - x)\parens{x - \half}^{k - 1} = (1 - x)f'(1 - x),
\end{align*}
as desired.
\end{proof}

\begin{lemma} \label{lem:constant}
The only infinitely differentiable scoring rules that are weakly proper but not proper are constant functions.
\end{lemma}

\begin{proof}[Proof of Lemma~\ref{lem:constant}]
Let $f$ be an infinitely differentiable scoring rule that is weakly proper but not proper. Recall that in the proof of Lemma~\ref{lem:weaklyproper}, we showed that for all $p$, the associated reward function $r_p(x)$ weakly increases on $(0, p]$ and weakly decreases on $[p, 1)$. Since $f$ is not proper, there is some $p$ such that $r_p(x)$ does not strictly increase on $(0, p]$ or does not strictly decrease on $[p, 1)$. But this means that $r_p(x)$ is constant on some open interval, which means that $r_p(x)$ is constant (because $f$ is infinitely differentiable, which means that $r_p(x)$ is also infinitely differentiable). Thus, for some $c \in \RR$ we have that $pf(x) + (1 - p)f(1 - x) = c$ for all $x$. Taking the derivative, we have that $pf'(x) = (1 - p)f'(1 - x)$ for all $x$. But we also have that $xf'(x) = (1 - x)f'(1 - x)$ for all $x$. The only way for both of these equations to hold is for $f'(x)$ to be uniformly zero, so $f$ is indeed constant.
\end{proof}

In proving Theorem~\ref{thm:approx}, we will make substantial use of $R''$, the second derivative of the reward function of $f$. For convenience, we will write $\phi$ instead of $R''$.
\begin{proof}[Proof of Lemma~\ref{lem:taylorphil}]

By Theorem~\ref{thm:analytic}, $f$ is a proper scoring rule if and only if $f$ is nonconstant, $f'(x) \ge 0$ everywhere, and
\[f(x) = c_0 + \sum_{k > 0 \text{\emph{ odd}}} c_k(2k + 1 - 2kx)\parens{x - \half}^k\]
for some $c_0, c_1, c_3, c_5, \dots \in \RR$. Equivalently,
\begin{align*}
f'(x) &= \sum_{k > 0 \text{\emph{ odd}}} kc_k(2k + 1 - 2kx)\parens{x - \half}^{k - 1} - 2kc_k\parens{x - \half}^k\\
&= \sum_{k > 0 \text{\emph{ odd}}} kc_k\parens{2k + 1 - 2kx - 2 \parens{x - \half}}\parens{x - \half}^{k - 1}\\
&= \sum_{k > 0 \text{\emph{ odd}}} 2kc_k(k + 1)(1 - x)\parens{x - \half}^{k - 1}\\
&= \sum_{k \ge 0 \text{\emph{ even}}} 2(k + 1)c_{k + 1}(k + 2)(1 - x)\parens{x - \half}^k = (1 - x) \sum_{k \ge 0 \text{\emph{ even}}} d_k \parens{x - \half}^k,
\end{align*}
where $d_k = 2(k + 1)(k + 2)c_{k + 1}$. Noting that $f$ is constant if and only if $\phi$ is uniformly zero, and that $f'(x) \ge 0$ if and only if $\phi(x) \ge 0$, this completes the proof.
\end{proof}

Note that if $f$ is a continuously differentiable (but not necessarily infinitely differentiable) proper scoring rule, then we have
\[\ind^\ell(f) = 2 \int_\half^1 \parens{\frac{x(1 - x)}{\phi(x)}}^{\ell/4} dx.\]

Also, note that $f$ is normalized if and only if $f \parens{\half} = 0$ and (by Corollary~\ref{cor:normalized})
\[\int_\half^1 (1 - x)f'(x) dx = \int_\half^1 (1 - x)^2\phi(x) dx = 1.\]

We now prove Theorem~\ref{thm:approx}.

\begin{proof}[Proof of Theorem~\ref{thm:approx}]

The Weierstrass approximation theorem says that any continuous function can be uniformly approximated by polynomials on a closed interval. A constructive proof of this theorem (for the interval $[0, 1]$) is given by the Bernstein polynomials: $b_{i, n}(x) = \binom{n}{i}x^i(1 - x)^{n - i}$. Given a continuous function $\psi: [0, 1] \to \RR$, define
\[B_n(\psi)(x) = \sum_{i = 0}^n \psi \parens{\frac{i}{n}} b_{i, n}(x).\]
Then the polynomials $B_n(\psi)$ converge uniformly to $\psi$ \cite[\S36]{Estep}. Suppose that $\psi$ also satisfies $\psi(x) = \psi(1 - x)$. Then $\psi \parens{\frac{i}{n}} = \psi \parens{\frac{n - i}{n}}$, which means $B_n(\psi)$ can be written as a linear combination of polynomials $(b_{i, n} + b_{n - i, n})(x)$. These polynomials are equal at $x$ and $1 - x$, and thus $B_n(\psi)(x) = B_n(\psi)(1 - x)$. From this we conclude that $\psi$ can be uniformly approximated on $[0, 1]$ by a sequence polynomials $p_i$ that satisfy $p_i(x) = p_i(1 - x)$.

Let $\phi_\ell$ be the $\phi$ corresponding to $g_{\ell, \opt}$. Recall that
\begin{equation} \label{eq:phi_ell}
\phi_\ell(x) = \begin{cases}\kappa_\ell (x^{\ell - 8}(1 - x)^\ell)^{1/(\ell + 4)} & x \le \half \\
\kappa_\ell (x^\ell(1 - x)^{\ell - 8})^{1/(\ell + 4)} & x \ge \half.\end{cases}
\end{equation}

Let $0 < \epsilon < \min(\half, \phi_\ell(\half))$. Consider the following function $\phi_{\ell, \epsilon}: [0, 1] \to \RR$.
\[\phi_{\ell, \epsilon}(x) = \begin{cases}\phi_\ell(\epsilon) & x \le \epsilon \\ \phi_\ell(x) & \epsilon \le x \le 1 - \epsilon \\ \phi_\ell(1 - \epsilon) & x \ge 1 - \epsilon.\end{cases}\]

Observe that $\phi_{\ell, \epsilon}(x) = \phi_{\ell, \epsilon}(1 - x)$ for all $x \in [0, 1]$; this is a straightforward consequence of the fact that $\phi_\ell$ is symmetric about $\half$. Per our discussion above, there exists a polynomial $p_\epsilon$ satisfying $p_\epsilon(x) = p_\epsilon(1 - x)$ such that for all $x \in [0, 1]$, $\abs{p_\epsilon(x) - \phi_{\ell, \epsilon}(x)} \le \epsilon$. In particular, we take $p_\epsilon = B_{n(\epsilon)}(\phi_{\ell, \epsilon})$, where $n(\epsilon)$ is any $n$ large enough that $p_\epsilon$ is uniformly within $\epsilon$ of $\phi_{\ell, \epsilon}$.

Observe that such a polynomial, when written as a sum of powers of $x - \half$, must only contain even powers of $x - \half$, since $p_\epsilon(x) - p_\epsilon(1 - x)$ must be the zero polynomial. Consequently, by Lemma~\ref{lem:taylorphil}, $(1 - x)p_\epsilon(x)$ is the derivative of a proper scoring rule.\footnote{The fact that $p_\epsilon$ is nonnegative everywhere follows from the fact that it is a uniform $\epsilon$-approximation of $\phi_{\ell, \epsilon}$, which is greater than $1$ on $[0, 1]$.} To find the associated normalized proper scoring rule (call it $f_\epsilon$), we take the antiderivative (taking the constant coefficient in the $\parens{x - \half}$-expansion to be zero), and divide by $\int_\half^1 (1 - x)^2p_\epsilon(x) dx$. Thus, corresponding to each $\epsilon$ we have a normalized polynomial scoring rule $f_\epsilon$ with incentivization index
\[\ind^\ell(f_\epsilon) = 2 \int_\half^1 \parens{\frac{x(1 - x)}{p_\epsilon(x) \cdot \frac{1}{\int_\half^1 (1 - x)^2p_\epsilon(x) dx}}}^{\ell/4} dx = 2 \parens{\int_\half^1 (1 - x)^2p_\epsilon(x) dx}^{\ell/4} \int_\half^1 \parens{\frac{x(1 - x)}{p_\epsilon(x)}}^{\ell/4} dx.\]

\begin{claim}
$f_\epsilon$ is respectful.
\end{claim}

\begin{proof}
Since $f_\epsilon$ is polynomial (and thus bounded and infinitely differentiable), it suffices to show that the second derivative of its reward function is bounded away from zero. The second derivative of $f_\epsilon$'s reward function is a positive multiple of $p_\epsilon$, so it suffices to show that $p_\epsilon$ is bounded away from zero. This is indeed the case. To see this, note that $\phi_{\ell, \epsilon}$ is bounded away from zero (as $\phi_\ell$ is bounded away from zero on $[\epsilon, 1 - \epsilon]$); let $L > 0$ be such that $\phi_{\ell, \epsilon}(x) \ge L$ on $[0, 1]$. Then
\[p_\epsilon(x) = \sum_{i = 0}^{n(\epsilon)} \phi_{\ell, \epsilon} \parens{\frac{i}{n(\epsilon)}} \binom{n(\epsilon)}{i} x^i (1 - x)^{n(\epsilon) - i} \ge \sum_{i = 0}^{n(\epsilon)} L \binom{n(\epsilon)}{i} x^i (1 - x)^{n(\epsilon) - i} = L.\]
\end{proof}

Our goal is to upper bound $\ind^\ell(f_\epsilon)$ in a way that shows that $\lim_{\epsilon \to 0} \ind^\ell(f_\epsilon) = \ind^\ell(g_{\ell,\opt})$. To do this, it suffices to show that the first of the two integrals in our formula for $f_\epsilon$ converges to $1$ as $\epsilon \to 0$ and that twice the second integral converges to $\ind^\ell(g_\ell, \opt)$ as $\epsilon \to 0$. We begin by working with the first of the two integrals.

\begin{claim}
\[\limsup_{\epsilon \to 0} \int_\half^1 (1 - x)^2p_\epsilon(x) dx \le 1.\]
\end{claim}

\begin{proof}
First observe that, since $g_{\ell,\opt}$ is normalized, we have $\int_\half^1 (1 - x)^2\phi_\ell(x) dx = 1$. Next, note that for $1 \le \ell \le 8$, $\phi_\ell$ is increasing on $[\half, 1)$ (as is evident from Equation~\ref{eq:phi_ell}), which means that
\[\int_\half^1 (1 - x)^2\phi_{\ell, \epsilon}(x) dx \le \int_\half^1 (1 - x)^2 \phi_\ell(x) dx = 1. \qquad (1 \le \ell \le 8)\]
On the other hand, for $\ell > 8$, observe that $\phi_\ell$ is bounded above on $[\half, 1)$, say by a constant $M_\ell$, which means that in this case
\[\int_\half^1 (1 - x)^2\phi_{\ell, \epsilon}(x) dx \le \int_\half^1 (1 - x)^2 \phi_\ell(x) dx + \epsilon M_\ell = 1 + \epsilon M_\ell. \qquad (\ell > 8)\]

Now, we have
\[\parens{\int_\half^1 (1 - x)^2p_\epsilon(x) dx}^{\ell/4} \le \parens{\int_\half^1 (1 - x)^2(\phi_{\ell, \epsilon}(x) + \epsilon) dx}^{\ell/4} \le \parens{\epsilon + \int_\half^1 (1 - x)^2\phi_{\ell, \epsilon}(x) dx}^{\ell/4}\]
which is at most $(1 + \epsilon)^{\ell/4}$ (for $\ell \le 8$) and at most $(1 + (M_\ell + 1)\epsilon)^{\ell/4}$ (for $\ell > 8$).
\end{proof}

Next we work with the second integral.

\begin{claim}
\[\limsup_{\epsilon \to 0} 2 \int_\half^1 \parens{\frac{x(1 - x)}{p_\epsilon(x)}}^{\ell/4} dx \le I(g_\ell, \opt).\]
\end{claim}

\begin{proof}
We have
\begin{align*}
\int_\half^1 \parens{\frac{x(1 - x)}{p_\epsilon(x)}}^{\ell/4} dx &\le \int_\half^1 \parens{\frac{x(1 - x)}{\phi_{\ell, \epsilon}(x) - \epsilon}}^{\ell/4} dx = \int_\half^1 \parens{\frac{x(1 - x)}{\phi_{\ell, \epsilon}(x)\parens{1 - \frac{\epsilon}{\phi_{\ell, \epsilon}(x)}}}}^{\ell/4} dx\\
&\le \parens{1 - \max_{x \in [\half, 1]} \frac{\epsilon}{\phi_{\ell, \epsilon}(x)}}^{\ell/4} \int_\half^1 \parens{\frac{x(1 - x)}{\phi_{\ell, \epsilon}(x)}}^{\ell/4} dx.
\end{align*}

Let us consider $\max_{x \in [\half, 1]} \frac{\epsilon}{\phi_{\ell, \epsilon}(x)}$ as a function of $\epsilon$. For $\ell \le 8$, since $\phi_\ell$ is increasing on $[\half, 1)$, this is just $\frac{\epsilon}{\phi_\ell(\half)}$, a quantity that approaches zero as $\epsilon$ approaches zero.

For $\ell > 8$, as $\epsilon$ approaches zero we have that $\min_{x \in [\half, 1]} \phi_{\ell, \epsilon}(x)$ approaches $\kappa_\ell \epsilon^{(\ell - 8)/(\ell + 4)} = \omega(\epsilon)$. This means that $\max_{x \in [\half, 1]} \frac{\epsilon}{\phi_{\ell, \epsilon}(x)}$ approaches zero as $\epsilon$ approaches zero.

Therefore, we have
\[\limsup_{\epsilon \to 0} 2 \int_\half^1 \parens{\frac{x(1 - x)}{p_\epsilon(x)}}^{\ell/4} dx \le \limsup_{\epsilon \to 0} 2 \int_\half^1 \parens{\frac{x(1 - x)}{\phi_{\ell, \epsilon}(x)}}^{\ell/4}.\]

Next, note that for $x \in [1 - \epsilon, 1]$, we have
\begin{align*}
\parens{\frac{x(1 - x)}{\phi_{\ell, \epsilon}(x)}}^{\ell/4} &= \parens{\frac{x(1 - x)}{\phi_\ell(1 - \epsilon)}}^{\ell/4} \le \parens{\frac{\epsilon(1 - \epsilon)}{\phi_\ell(1 - \epsilon)}}^{\ell/4} = \parens{\frac{\epsilon(1 - \epsilon)}{\kappa_\ell ((1 - \epsilon)^\ell \epsilon^{\ell - 8})^{1/(\ell + 4)}}}^{\ell/4}\\
&= \frac{(\epsilon^3(1 - \epsilon))^{\ell/(\ell + 4)}}{\kappa_\ell^{\ell/4}} \le \frac{1}{\kappa_\ell^{\ell/4}}.
\end{align*}
Therefore we have
\[\int_\half^1 \parens{\frac{x(1 - x)}{\phi_{\ell, \epsilon}(x)}}^{\ell/4} dx \le \int_\half^1 \parens{\frac{x(1 - x)}{\phi_\ell(x)}}^{\ell/4} dx + \frac{\epsilon}{\kappa_\ell^{\ell/4}} = \half \ind^\ell(g_{\ell,\opt}) + \frac{\epsilon}{\kappa_\ell^{\ell/4}},\]
so
\[\limsup_{\epsilon \to 0} 2 \int_\half^1 \parens{\frac{x(1 - x)}{\phi_{\ell, \epsilon}(x)}}^{\ell/4} \le \ind^\ell(g_{\ell, \opt}).\]
\end{proof}

It therefore follows that $\limsup_{\epsilon \to 0} \ind^\ell(f_\epsilon) \le \ind^\ell(g_{\ell,\opt})$. But in fact, the inequality is an equality; this is because no continuously differentiable function has incentivization index less than that of $g_{\ell,\opt}$. This completes the proof of Theorem~\ref{thm:approx}.
\end{proof}
\section{Simulation Results} \label{app:simulation}
The table below summarizes the results of a quick simulation for $\ell =1$. Note that ``predicted average error" simply means the error that Theorem~\ref{thm:global} predicts in the limit as $c$ approaches $0$ (but multiplied by $c^{-1/4}$ in the stated value of $c$). ``Ratio" refers to the ratio between the average error and the predicted average error. ``Maximum number of flips" refers to the maximum number of flips for that cost and rule in the $100,000$ simulations. (Note that because of the large number of simulations and comparatively small number of flips, these are likely to be universal upper bounds on the number of flips given that cost and rule.)

\small
\begin{center}
\begin{tabular}{|c c||c|c|c|c|c|}
\hline
Cost & Rule & Avg. Error & Predicted Avg. Error & Ratio & Avg. \# Flips & Max. \# Flips\\
\hline \hline
0.1 & $g_\text{quad}$ & 0.1616 & 0.1490 & 1.0845 & 2.3341 & 3\\
& $g_\text{log}$ & 0.1609 & 0.1389 & 1.1582 & 2.3317 & 3\\
& $g_{1,\opt}$ & 0.1553 & 0.1348 & 1.1522 & 2.6658 & 3\\
\hline
0.03 & $g_\text{quad}$ & 0.1136 & 0.1103 & 1.0298 & 6.0933 & 7\\
& $g_\text{log}$ & 0.1093 & 0.1028 & 1.0636 & 6.7133 & 7\\
& $g_{1,\opt}$ & 0.1110 & 0.0997 & 1.1126 & 7.1547 & 10\\
\hline
0.01 & $g_\text{quad}$ & 0.0850 & 0.0838 & 1.0147 & 11.9745 & 15\\
& $g_\text{log}$ & 0.0816 & 0.0781 & 1.0444 & 13.2780 & 14\\
& $g_{1,\opt}$ & 0.0802 & 0.0758 & 1.0577 & 15.3399 & 23\\
\hline
0.003 & $g_\text{quad}$ & 0.0626 & 0.0620 & 1.0096 & 23.2076 & 29\\
& $g_\text{log}$ & 0.0590 & 0.0578 & 1.0199 & 26.3918 & 27\\
& $g_{1,\opt}$ & 0.0580 & 0.0561 & 1.0349 & 31.2093 & 52\\
\hline
0.001 & $g_\text{quad}$ & 0.0472 & 0.0471 & 1.0014 & 41.5592 & 52\\
& $g_\text{log}$ & 0.0448 & 0.0439 & 1.0193 & 47.4845 & 48\\
& $g_{1,\opt}$ & 0.0434 & 0.0426 & 1.0179 & 57.5323 & 107\\
\hline
0.0003 & $g_\text{quad}$ & 0.0349 & 0.0349 & 1.0016 & 77.2931 & 97\\
& $g_\text{log}$ & 0.0329 & 0.0325 & 1.0113 & 89.2927 & 90\\
& $g_{1,\opt}$ & 0.0320 & 0.0315 & 1.0130 & 108.8362 & 230\\
\hline
0.0001 & $g_\text{quad}$ & 0.0265 & 0.0265 & 0.9999 & 134.6477 & 171\\
& $g_\text{log}$ & 0.0248 & 0.0247 & 1.0047 & 157.1403 & 158\\
& $g_{1,\opt}$ & 0.0241 & 0.0240 & 1.0043 & 192.5842 & 460\\
\hline
0.00003 & $g_\text{quad}$ & 0.0196 & 0.0196 & 1.0013 & 247.0952 & 314\\
& $g_\text{log}$ & 0.0184 & 0.0183 & 1.0056 & 289.8925 & 291\\
& $g_{1,\opt}$ & 0.0177 & 0.0177 & 0.9980 & 356.4421 & 979\\
\hline
\end{tabular}
\end{center}
\end{document}